\documentclass[aps,amsmath,prd, showpacs]{revtex4}

\usepackage{graphicx}

\def\l{\left}
\def\r{\right}
\def\be{\begin{equation}}
\def\ee{\end{equation}}
\def\bea{\begin{eqnarray}}
\def\eea{\end{eqnarray}}

\begin{document} \title{
Searching for Gravitational Waves from the Inspiral of Precessing Binary
Systems: New Hierarchical Scheme using ``Spiky'' Templates}

\author{Philippe Grandcl\'ement}
\email[]{PGrandclement@northwestern.edu}
\affiliation{Northwestern University, Dept.\ of Physics \& Astronomy, 2145 
Sheridan Road, Evanston 60208, USA}

\author{Vassiliki Kalogera}
\email[]{vicky@northwestern.edu}
\affiliation{Northwestern University, Dept.\ of Physics \& Astronomy, 2145 
Sheridan Road, Evanston 60208, USA}

\date{January 20, 2002}

 \begin{abstract}

 In a recent investigation \cite{GrandKV02} of the effects of precession
on the anticipated detection of gravitational-wave inspiral signals from
compact object binaries with moderate total masses $\lesssim
10\,M_{\odot}$, we found that (i) if precession is ignored, the inspiral
detection rate can decrease by almost a factor of $10$, and (ii)  
previously proposed \cite{Apost96} ``mimic'' templates cannot improve the
detection rate significantly (by more than a factor of 2). In this paper
we propose a new family of templates that can improve the detection rate
by factors of 5--6 in cases where precession is most important. Our
proposed method for these new ``mimic'' templates involves a hierarchical
scheme of efficient, two-parameter template searches that can account for
a sequence of spikes that appear in the residual inspiral phase, after one
corrects for the any oscillatory modification in the phase.
We present our results for two cases of compact object masses
($10$ and $1.4$\,M$_\odot$ and $7$ and $3$\,M$_\odot$) as a function of
spin properties. Although further work is needed to fully assess the
computational efficiency of this newly proposed template family, we
conclude that these ``spiky templates'' are good candidates for a family
of precession templates used in realistic searches, that can improve
detection rates of inspiral events.
 \end{abstract}

\pacs{04.80.Nn  95.75.-z  95.85.Sz  97.60.-s}
\maketitle

 \section{Introduction}\label{s:intro}
 
 The inspiral of binary compact objects is one of the main targets of the
ground-based, interferometric gravitational-wave detectors currently
coming online (LIGO \cite{Abram92}, VIRGO \cite{Caron97}, GEO600
\cite{Danzm95} and TAMA \cite{Tagos01}). It is well known that the most
effective way of extracting such signals from the intrinsic noise of the
detectors is to use {\em matched-filtering} techniques
\cite{Helst68,OwenS99,Finn99}, which involves the correlation of detector
output with a family of templates thought to represent the expected
signals. It is clear that the success of such searches critically depends
on the accuracy of the adopted template families.

A great effort has been devoted to the computation of compact object
inspiral waveforms, especially using various PN-expansions (see
\cite{CutleT02} for a review). The current searches in LIGO, GEO600 and
TAMA, are performed using the template family corresponding to two,
non-spinning, point masses, including 2.5-PN order corrections. However,
in some cases, expected signals are believed to deviate significantly from
these waveforms. For the case of massive compact binaries (total mass in
excess of $10$\,M$_\odot$), it has been shown that the PN approximations
break down right in the frequency band of the interferometric detectors
\cite{BradyCT98}. In fact the exact signal is unknown and a number of
different approximations have been suggested as alternatives.  Recently, a
family of templates that exhibits good overlap with the results obtained
using all the various expansion approximations has been suggested for use
in future searches (see \cite{BuonaCV02} and references therein).

Another case where the current template family does not reflect realistic
signals is the case of {\em precessing} binaries. Compact objects with
spins of significant magnitude and misalignment relative to the orbital
angular momentum axis emit inspiral signals that are mainly modified by
precession of the orbital plane \cite{ApostCST94}. This precession is
caused by high-order spin-orbit and spin-spin couplings. Depending on the
physical properties of the binary, a significant number of precession
cycles occur within the frequency band of current ground-based
interferometers. The resulting change in the polarization of the wave
leads to modulations of both the amplitude and the phase of the inspiral
signal. In principle, one would like to build a family of precessing
inspiral templates.  However, this is unrealistic because of computational
limitations:  precessing waveforms depend on extra parameters (spin
magnitudes and orientations) and make the dimension of template parameter
space prohibitively large. The first investigations of the importance of
precession \cite{ApostCST94} were further expanded in a consistent way
\cite{GrandKV02} (hereafter Paper I)  to account for the current LIGO
noise curve and all possible physical configurations of intermediate-mass
$\lesssim 10\,M_{\odot}$ systems. It is now understood that ignoring
precession in the templates could risk the anticipated inspiral detection,
since the detection rate can be reduced by almost an order of magnitude in
the worst cases. The need for a family of ``mimic'' templates that can
capture the signal modifications without unreasonably expanding the
dimensionality of the templates is clear. Early on \cite{Apost96}
suggested such a family that depends on only 3 additional parameters. In
Paper I we tested it extensively and found that this family alone did not
improve the detection rate significantly (improvement factors remained
lower than 2 regardless of spin properties and masses).

Our motivation for this paper is to pursue the issue of ``mimic''
precession templates further. Throughout this work the methods and
notations are the same as in Paper I.  The precessing signals are obtained
using the simple-precession formalism \cite{ApostCST94,Apost95}, where
only the most massive object carries a spin (see the Introduction of Paper
I for a justification). In this regime, the effect of precession is described by
both a phase and an amplitude modulation, so that, in the frequency 
domain, the signal is given by
\be\label{e:signal}
\tilde{h}_{\rm prec}\l(f\r) = {\rm AM} \times {\rm PM} \times
\tilde{h}_{\rm no \, prec}\l(f\r).
\ee
In Eq. (\ref{e:signal}), $\tilde{h}_{\rm no \, prec}$ denotes the signal with the 
same physical parameters (masses, orientations ...) but without precession.
${\rm AM}$ is an amplitude modulation (Eq. (11) of \cite{Apost95}) and 
${\rm PM} = \exp\l(-i\phi_{\rm mod}\r)$ is a phase modulation (Eq. (12) of 
\cite{Apost95}). We will neglect the Thomas precession (Eq. (14) of 
\cite{Apost95}), assuming that the monotonic modulation induced by 
it will not greatly influence the results. This assumption should be checked by
future work.
Except when otherwise stated, the {\em
non-precessing} parts of the templates and the signal,
$\tilde{h}_{\rm no \, prec}$,
are Newtonian. This
is a simplifying assumption that makes our calculations feasible, given
our current computational resources. We have already shown in Paper I that
this assumption does not affect our results and conclusion in any
quantitative way.

 In this paper we have again included some indicative
runs (of high computational cost) to show that the assumption of Newtonian
{\em non-precessing} parts of the templates and the signal (as long as the
two are consistent with each other; cf.\ \cite{Apost96}) is not in any way
limiting. We adopt a noise curve relevant to the initial LIGO. The
efficiency of a family of templates is quantified by the concept of the
{\em fitting factor} (FF) \cite{ApostCST94}, which describes the loss in
signal-to-noise ratio (SNR)  due to the mismatch between the signal and
the templates:
 \be
 \l(\frac{S}{N}\r) = \mathrm{FF} \times \l(\frac{S}{N}\r)_\mathrm{max}
 \ee
 where $\l(S/N\r)_\mathrm{max}$ is the SNR that would be achieved if the
family of templates included the signal (see
\cite{Apost96,Helst68,ApostCST94,Finn99} for more details).
Most of the results will be presented by plotting the quantity
$\mathrm{<FF>}^3$, the factor by which the detection rate is reduced,
assuming a uniform distribution of sources in volume and using the fitting
factor averaged over all random angles in the problem.

This paper is organized as follows. In Sec.\ref{s:new_family}, we first
provide a physical explanation for why the Apostolatos' ``mimic'' template
family is insufficient, in the context of an example case analyzed in
detail.  Motivated by the results in the first part, we introduce a new
family of templates, and show how they can improve the SNR of detections.  
After briefly reviewing the determination of various computational
parameters, we present our comprehensive results for two cases of high
precession ($\l(m_1 , m_2\r) = \l(10 , 1.4\r)$ and $\l(m_1 , m_2\r) = \l(7
, 3\r)$ in M$_\odot$). Conclusions, perspectives, and future work are
discussed in Sec.\ \ref{s:conclu}.

 \section{New family of templates}\label{s:new_family}

 \subsection{Plausible explanation for the insufficiency of Apostolatos' 
``mimic'' templates}\label{ss:explanation}

The Apostolatos' ``mimic'' templates consist of non-precessing waveforms 
modulated in phase by the following oscillatory mathematical form: 
 \be
 \label{e:Apost} 
 \phi^{\rm cor} = {\mathcal C} \cos \l({\mathcal B} f^{-2/3} + \delta\r) 
 \ee
 In the simple precession regime \cite{Apost96,ApostCST94,Apost95}, both
the spin $\vec{S}$ and the orbital momentum $\vec{L}$ precess around the
total angular momentum $\vec{J}$, which remains at an approximately
constant direction. Analytical, approximate formulae (Eqs. (29) of
\cite{Apost95}) of the precession angle
(i.e. the angle describing the position of $\vec{L}$ along its precession
cone), as a function of frequency, 
have been derived. It has been argued that the modulation of the
inspiral phase induced by precession would have essentially the same
behavior with respect to frequency, hence the form (\ref{e:Apost}).

There was also an expectation that any monotonic modulation of the phase
would be indirectly accounted for by a mismatch of the non-precessing
parameters (e.g., masses) with respect to the real values responsible for
the signal. In Paper I we have shown that this oscillatory modulation
alone cannot recover a high SNR (we have also shown that this conclusion
does not depend on the choice of the power index for the dependence of the
precession angle on frequency). As a first step in the search for a more
efficient family of templates, it is useful to try and understand in some
more detail the behavior of the Apostolatos' phase correction.

Before showing some particular examples, let us note some important
points. First of all, one of the effects of simple precession is an 
additive phase modification of the non-precession phase: 
 \be
\label{e:phimod}
 \phi_S = \phi_{\rm no \, prec} + \phi_{\rm mod}
 \ee
 where $\phi_S$ is the total phase of the signal, $\phi_{\rm no \, prec}$
the phase without precession, and $\phi_{\rm mod}$ the {\em phase
modulation}. The expression proposed by Apostolatos (\ref{e:Apost}) was
intended to ``mimic'' the true $\phi_{\rm mod}$.  However we will see that
this is not always the case.

Another complication comes from the fact that $\phi_{\rm mod}$ is not the
phase that one would like to ``mimic'' to obtain high SNRs, at least not
in the framework proposed by \cite{Apost96} and tested in Paper I. Indeed,
in these studies, searches are performed in two consecutive steps. First,
one uses only the (post)-Newtonian templates without any precession
modification. Let us call $T_{\rm max}$ the Newtonian template for which
the maximum ${\rm FF}$ is obtained, and ${\mathcal M}_{\rm max}$, $t_{\rm
max}$ and $\phi_{\rm max}$ the parameters of this template (the dependence
on $\phi_{\rm max}$ is analytical but its value can be determined, mainly
for plotting purposes). Because of the modulations of the signal induced
by precession, the parameters of $T_{\rm max}$ do not match exactly the
ones of the signal. Based on the expectation that the parameters of the
Newtonian templates do not correlate with those of the phase correction
(\ref{e:Apost}), the maximization over the parameters of this correction is
performed after $T_{\rm max}$ is identified.

However, because of the mismatch of the parameters, the phase difference
between $T_{\rm max}$ and the signal $S$ is {not} the phase modulation
$\phi_{\rm mod}$. Instead it is equal to what we will call the {\em 
residual} phase: 
 \be
\label{e:phires}
 \Delta\phi = \phi_S - \phi_{T_{\rm max}},
 \ee
 where $\phi_S$ is the phase of the signal and $\phi_{T_{\rm max}}$ the
phase of $T_{\rm max}$. In a 2-step search scheme (first Newtonian then
``mimic'' templates), the ``mimic'' templates must represent a good fit to
the residual phase defined above. It turns out that the residual phase
$\Delta\phi$ and the phase correction (\ref{e:Apost}) can have very
different behaviors so that this effect of the mismatch is rather crucial.

In what follows we consider two opposite examples of precession signals
and their behaviors. In both cases the values of the masses are $m_1 =
10$\,M$_\odot$ and $m_2 = 1.4$\,M$_\odot$, the spin magnitude of the most
massive compact object is maximum ($S=1$), and the cosine of the spin-tilt
angle (relative to the orbital angular momentum axis) is $\kappa \equiv
\vec{S} \cdot \vec{L} = 0.4$. Depending on the values of the random angles
that determine the orientation of the orbital plane ($\theta'$ and
$\varphi'$), the sky position ($\theta$ and $\varphi$), and the constant
phase present in the expression for the precession angle (cf.  Eq. (63a)
of \cite{ApostCST94}) ($\alpha_{\rm prec}$), the behavior of the phase 
modulation $\phi_{\rm mod}$  (Eq. (\ref{e:phimod})) can be very
different. One of the two examples (configuration I) produces a monotonic
modulation whereas the other (II) produces mainly an oscillation. The
exact values of the random angles, for both configurations are given in
Table \ref{t:config}.

 \begin{table}[h]
 \caption{\label{t:config} Values of the random angles for the two 
configurations shown respectively in Fig. \ref{f:mono} and \ref{f:oscille}}
 
 \begin{tabular}{|c||c|c|c|c|c|c|}
 \hline
  & Behavior & $\cos \theta$ & $\varphi$ & $\cos \theta'$ & $\varphi'$ &
$\alpha_{\rm prec}$ \\
 \hline
 Configuration I & Monotonic $\l[2\pi\r]$ & $-0.615$ & $1.31$ & $0.248$ & 
$3.86$ &  $3.32$ \\
 Configuration II & Oscillatory & $0.271$ & $3.67$ & $0.203$ & 
$1.48$ & $5.41$ \\
 \hline
 \end{tabular}
 \end{table}

First we focus on Configuration I (Figure 1). Using the signal parameters
we can calculate exactly $\phi_{\rm mod}$ and plot it as a function a
function of frequency (dashed line on left panel).  As already mentioned,
it is a rather monotonic function (recall that the plot shows the phase
modulo $2\pi$). We can also take the ``mimic'' templates, fix the
non-precessing parameters to be exactly equal to those of the signal
(i.e., we avoid any mismatch effects), and calculate the parameters of the
phase correction (\ref{e:Apost}) that maximize FF. We can then plot this
phase correction as a function of frequency (solid line on left panel).  
As expected, the oscillatory behavior of Eq. (\ref{e:Apost})  does not
match well the phase modulation. In reality we cannot fix the
non-precessing parameters of the template to be equal to those of the
signal. Instead a 2-step maximization process has been suggested (assuming
that the non-precessing parameters do not correlate with those of the
precession modulation). In this case, it was expected that the first
maximization (and subsequent mismatch of the non-precessing parameters)
would match the the monotonic behavior of the signal, and the second
maximization would match the oscillatory behavior. If this were the case,
then the residual phase $\Delta \phi$ (dashed line on right panel) would
be described well by $\phi^{\rm cor}$ (\ref{e:Apost}; solid line on right
panel). Comparison of these two curves shows that this expectation is not
realistic.  The monotonic behavior, modulo $2\pi$,
is still present. Thus the correction
(\ref{e:Apost}) (solid line of the right panel) does a really poor job.

 \begin{figure}
 \includegraphics[height=6.5cm]{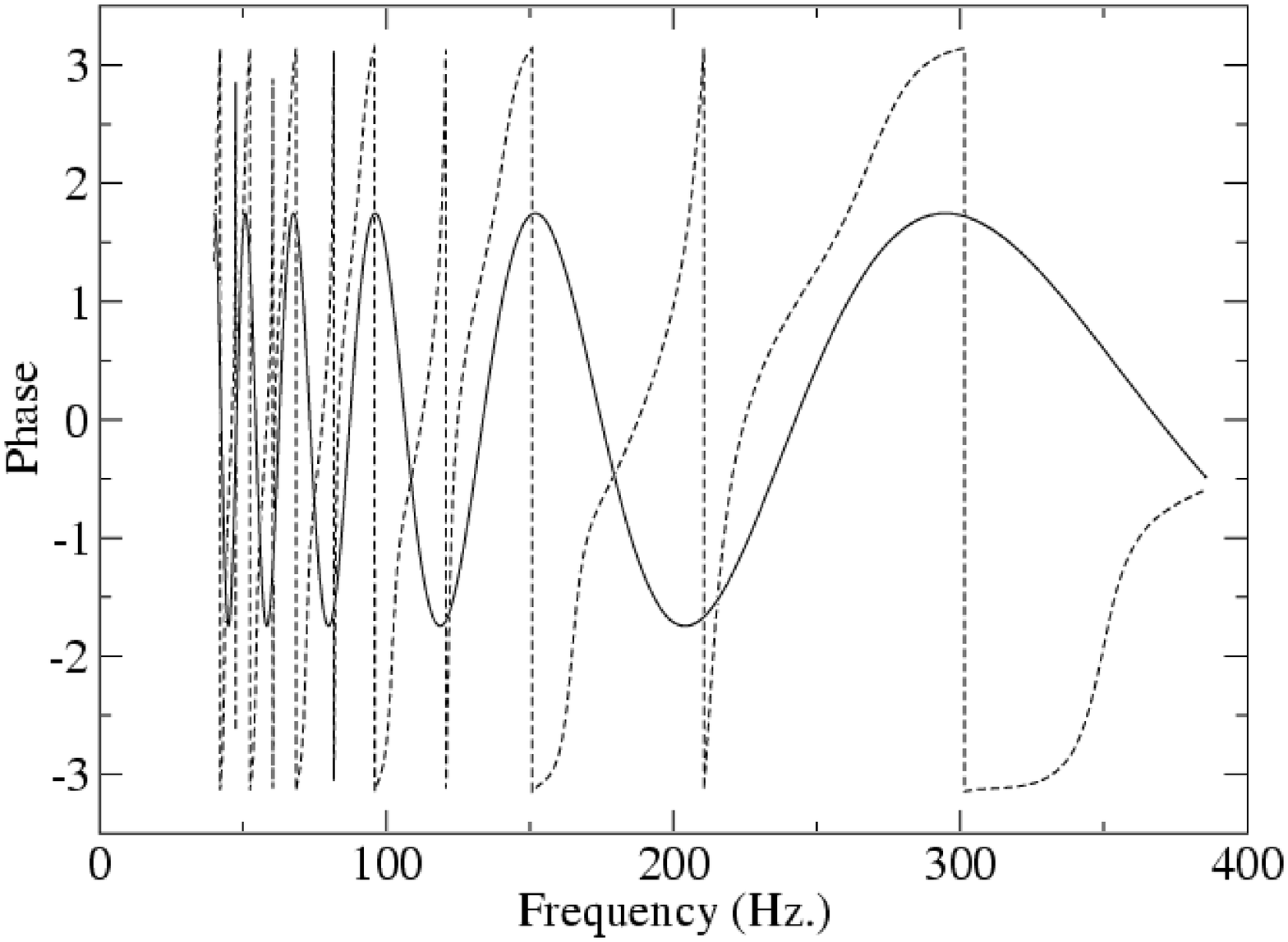} 
 \includegraphics[height=6.5cm]{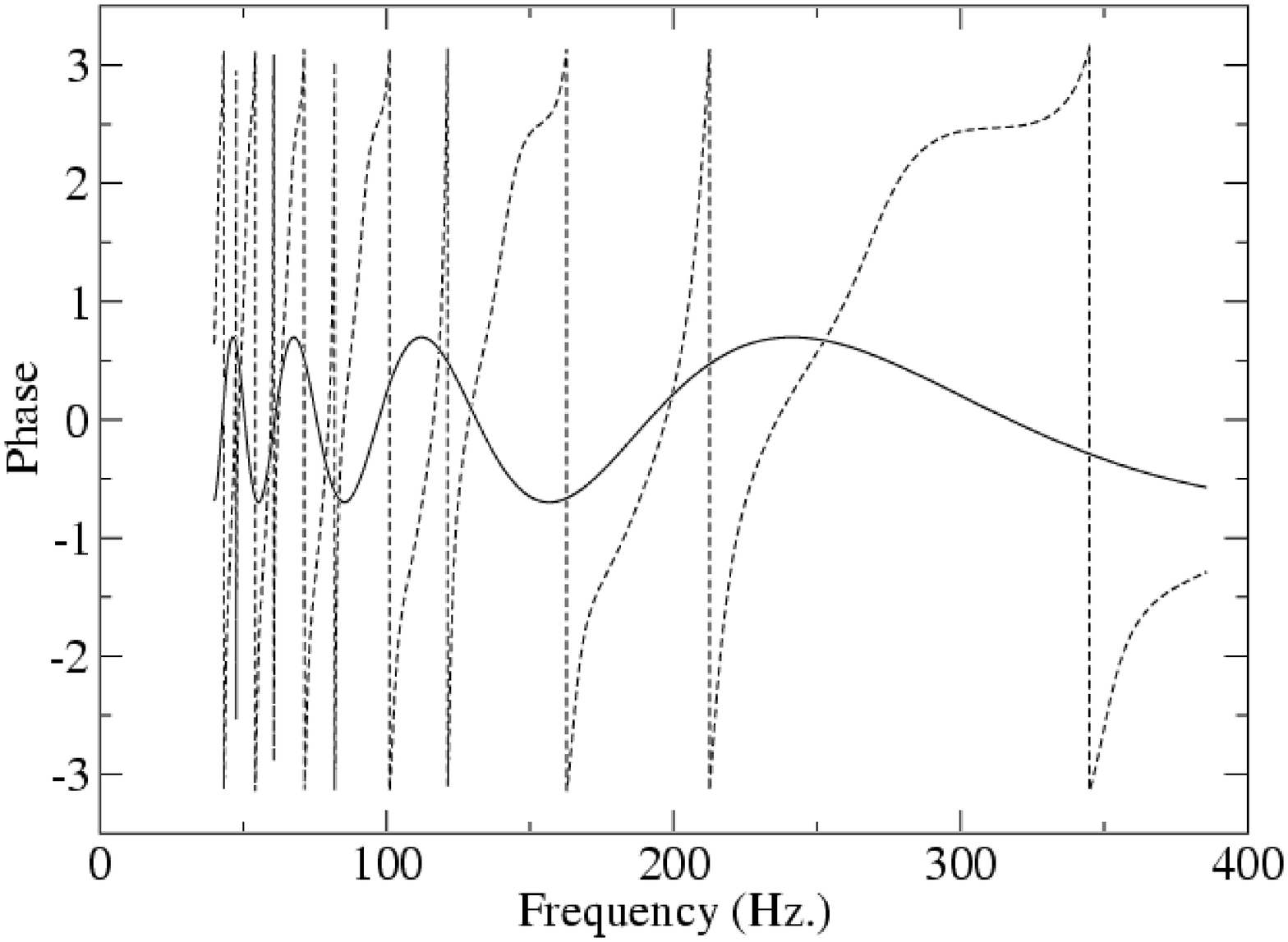}
 \caption{\label{f:mono} 
 The dashed line on the left panel shows the phase modulation $\phi_{\rm
mod}$ of the signal (Eq. (\ref{e:phimod})). The dashed line on the right
panel shows the residual phase $\Delta \phi$ (Eq. (\ref{e:phires})) after
the first maximization over non-precessing parameters of the templates.
The solid lines correspond to the best-fitting phase correction $\phi^{\rm
cor}$ (Eq. (\ref{e:Apost})) when the non-precessing parameters are set equal to
those of the signal (left panel) and when they are derived through a
maximization process, leading to a mismatch (right panel). For more
details see text.  The physical configuration is $m_1 = 10 M_\odot$, $m_2
= 1.4 M_\odot$, $S=1$ and $\kappa = 0.4$. The random angles are given by
the first line of Table \ref{t:config} (Configuration I).}
 \end{figure}

\begin{figure}
 \includegraphics[height=6.5cm]{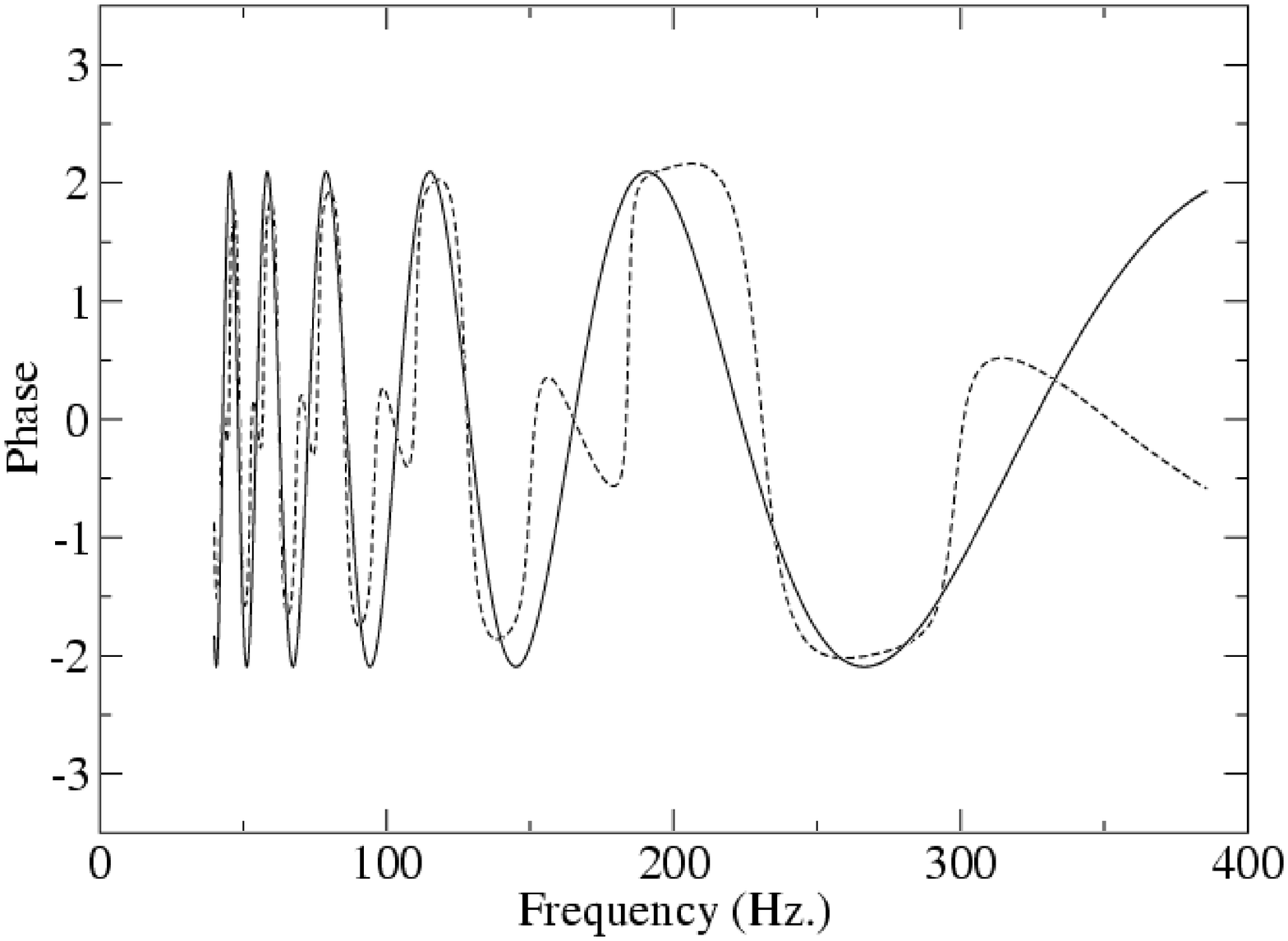} 
 \includegraphics[height=6.5cm]{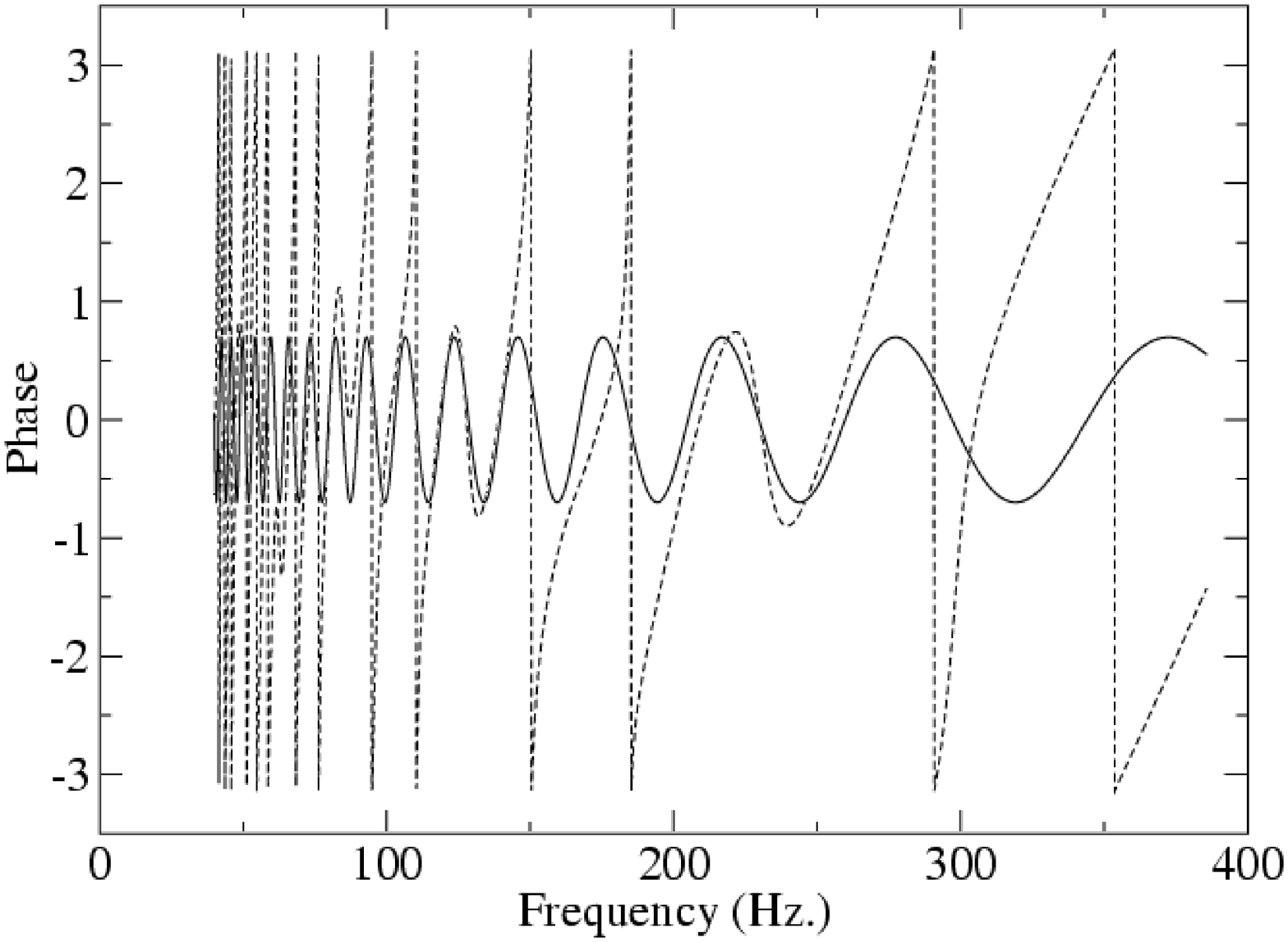}
 \caption{\label{f:oscille} Same as Fig. \ref{f:mono} by for Configuration 
II. }
 \end{figure}

The second configuration (II) differs qualitatively from the first one and
exhibits clear oscillatory behavior in the phase modulation $\phi_{\rm mod}$ 
itself (Eq. (\ref{e:phimod})) (dashed line of the left panel of Figure
\ref{f:oscille}). We can perform the same tests as with Configuration I.
We fix the Newtonian parameters of the templates to those of the signal,
and maximize over the three parameters of (\ref{e:Apost}). It turns out
that in this case the fit is quite good, and this maximization raises the
FF from $0.21$ to $0.67$. However, in a real search, the non-precessing
parameters are unknown, as we noted before. It turns out that the first
maximization over the non-precessing parameters introduces a monotonic
behavior ($\l[2\pi\r]$) to the residual phase $\Delta \phi$ (Eq. (\ref{e:phires}))
(dashed line on right panel). As a result the second maximization (this
time over the three parameters of (\ref{e:Apost}) does not lead to a high
improvement of the ${\rm FF}$: from $0.55$ after the use of of the
Newtonian templates it increases to to $0.67$ after the addition of
(\ref{e:Apost}). The fact that the two final ${\rm FF}$ are almost equal is 
fortuitous. Indeed, in \cite{Apost96}, Apostolatos made the assumption that 
the non-precessing variables ${\mathcal M}$ and $t_c$ are not correlated 
to the parameters of the correction (\ref{e:Apost}). If this is the case, 
the ${\rm FF}$ obtained by doing the maximization in two steps is the same than
the one that would be computed doing a five-dimensional search. The ${\rm FF}$ 
obtained by fixing the Newtonian parameters could only be smaller (or equal in this
extreme case). From our computation, it seems that this uncorrelation is 
not as good as stated in \cite{Apost96}. However, this needs further exploration, 
beyond the scope of this paper.

The two previous examples illustrate the fact that the oscillatory
correction proposed by Apostolatos does not recover the main part of the
signal because the residual phase $\Delta\phi$ exhibits some kind of
monotonic behavior, which is either inherent to the precession phase
modulation (configuration I) or is introduced after the first-step
maximization over the non-precessing parameters (configuration II). We have
performed many more example studies like the above with consistent
results, but of course this analysis does not constitute a proof. Given
the very complicated dependence of the precession phase on a large number
of parameters (some non-physical, e.g., orientation angles), it is
difficult to imagine how one can derive an exact proof of the qualitative
characteristics of the precessing inspiral signal. However, these example
studies provided us with motivation for the development and the choice of
a new template family that has been tested, increases FF values, and is
presented next.

 \subsection{Spikes in the phase residuals}\label{ss:spikes}

\begin{figure} 
 \includegraphics[height=10cm]{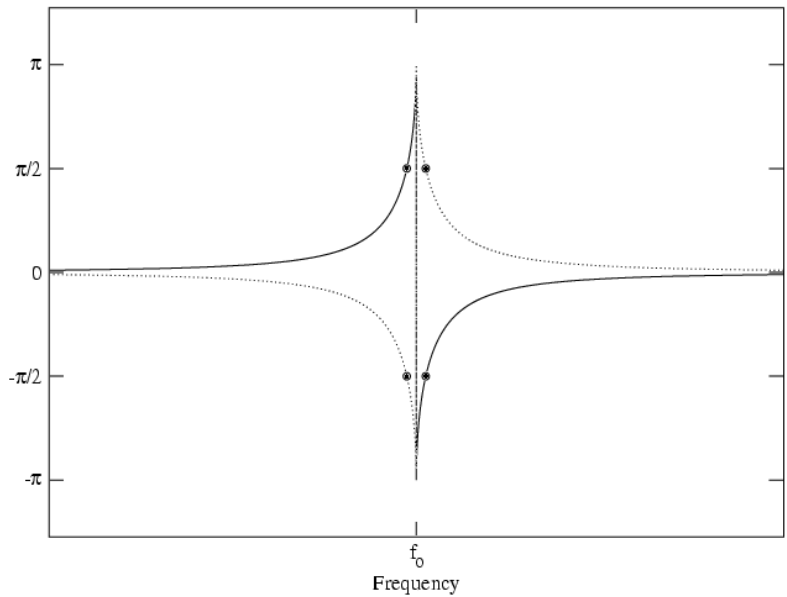}
 \caption{\label{f:spike} Function $P$. The solid line corresponds to 
$\varepsilon = 1$ and the dotted line to $\varepsilon = -1$. The
central position is $f_0$. A measure of the width of the spike is given by
the black dots are located at half width, i.e. at $f_0 \pm f_{1/2}$.}
 \end{figure}

Our extensive study of specific precession waveforms helped us realize
that the main features of Figures \ref{f:mono} and \ref{f:oscille} are
rather common.  Indeed the monotonic behavior can be represented by a
succession of spike-like structures centered further apart as frequency
increases.  The origin of the spikes in this representation of the phase
(modulo $2\pi$) is related to the fact that the residual phase jumps from
$0$ to $\pi$ and $0$ to $-\pi$ on either side of each spike. To reflect
this rapid change in phase we choose a mathematical form with an infinite
derivative at the central frequency of the spike $f_0$:
 \bea
 \label{e:spike}
 \mathrm{If} \quad f>f_0 \quad  \mathrm{then} \quad P\l(f_0, \sigma, 
\varepsilon\r) 
 &=& \varepsilon \pi \l[\sqrt{\l(1-\frac{1}
{\l(\sigma\left(f-f_0\r)+1\r)^2}\r)}-1\r] \\
 \nonumber
 \mathrm{If} \quad f<f_0 \quad  \mathrm{then} \quad P\l(f_0, \sigma, 
\varepsilon\r) 
 &=& \varepsilon \pi \l[-\sqrt{\l(1-\frac{1}{\l(
\sigma\l(f-f_0\r)-1\r)^2}\r)}+1\r].
 \eea
 The functions $P$ (Fig. \ref{f:spike}) 
depend on three well constrained parameters :
 \begin{itemize}
 \item $\varepsilon$ is either $1$ or $-1$ and represents the two possible 
orientations of the spike.
 \item $\sigma$ is related to the width of the spike. More precisely the 
half width is $f_{1/2} = \l(2\sqrt{3}-3\r)/\l(3\sigma\r)$.
 \item $f_0$ is the central position of the spike.
 \end{itemize}

To recover most of the residual phase $\Delta \phi$ (Eq. (\ref{e:phires}))
(after the first-step maximization over non-precessing parameters has been
performed), like those in the right panels of Figures \ref{f:mono} and
\ref{f:oscille}, we need to account for a sequence of a few or several
spikes. By definition, the functions $P$ represent narrow spikes, and
hence their value is non-zero only very close to $f_0$. Thus, a sum of
functions $P$, with a range of $f_0$, $\sigma$, and $\varepsilon$ values
should give us a good representation of the qualitative behavior of
residual phase with frequency. In principle, a prescription providing us
with the number of spikes, their positions, and their width is needed. The
main constraints are that (i) such prescriptions must depend on as few
parameters as possible and (ii) they must recover the number and
properties of spikes well enough to increase FF significantly.  We tested
various possible choices, but no accurate, simple prescription was found.  
The reason is rather obvious given the morphology of these features of the
residual phase: the functions $P$ are strongly localized in frequency
space, so that the positions $f_0$ must coincide very precisely with the
location of the spikes in the residual phase.

If the problem is related to the fact that $P$ is a highly localized
function, so does the solution. Indeed, given their behavior, it seems
natural to assume that the various spikes do not influence each other very
much. More precisely, the correction of one particular spike will not
influence the properties of the others, precisely because the functions
$P$ are non-zero only very close to their center frequency. This leads us
to consider a hierarchical search, where spike after spike is identified
sequentially. Each step in this sequence involves just a 2-parameter
search ($f_0$ and $\sigma$; $\varepsilon$ only has two possible values and
therefore can be taken into account explicitly by doubling the extent of
the search in the other two parameters). We note that $\sigma$ is narrowly
constrained, since the spikes of relevance are always very narrow (highly
localized). Two-parameter searches are repeated until the relative change
in FF drops below a certain threshold, typically $10^{-2}$.

However, we found that, in some cases, an oscillatory behavior can be
present (e.g., configuration II in the previous subsection). Therefore it
seems reasonable to keep the Apostolatos phase correction (\ref{e:Apost}).
Such maximization can ``mimic'' accurately most of oscillations in the
residual phase (note, for example, the two bumps on the right panel of
Fig. \ref{f:oscille}, around 125 and 220 Hz). Therefore a hierarchical set of
the 2-parameter searches for the spikes must follow.

Depending on the behavior of the residual phase, three types of situations
can be expected :
 \begin{itemize} 
 \item The residual phase is oscillatory. In such a case a phase
correction of the form (\ref{e:Apost}) recovers most part of the signal
and ${\rm FF}$ increases. Not many spikes, if any, are expected, and
subsequent searches with the ``spiky'' templates do not improve ${\rm FF}$
(it is important to note that no computational effort is wasted with the
spiky template searches, since those are aborted if the increase in ${\rm
FF}$ is not significant).
 \item The residual phase contains both spikes and oscillations (e.g.,
right panel of Figure \ref{f:oscille}). Phase corrections (\ref{e:Apost})
will trace the oscillations, and only moderately ``pollute'' the
spikes. The hierarchical spike search is necessary and the ${\rm FF}$
increases substantially after both types of maximizations.
 \item The residual phase
contains only spikes. The oscillator (\ref{e:Apost}) does a really poor
job (e.g., right panel of Figure \ref{f:mono}) in recovering the 
signal-to-noise ratio. However, this first search for oscillatory 
corrections does not introduce significant spurious traces to the residual 
phase. Several spikes are found and lead to a significant increase of 
${\rm FF}$. 
 \end{itemize} 
 These three types are rather simplistic in their description, but we have
found that they cover the full range of possible situation at roughly
equal weights, for the full range of random angles that determine the
signal as ``seen'' by the detector. 

Let us finally comment on the role of the oscillatory phase correction
(\ref{e:Apost}). It was initially motivated as a correction that will
closely track the number of precession cycles (Eqs. (29) of
\cite{Apost95}). If this were true, the pulsation of the oscillation
(expressed by ${\mathcal B}$) should lie close to the one given by Eqs.
(29) of \cite{Apost95}.  
However, we found in Paper I that this is not the case, and that the
values of ${\mathcal B}$ for which the maximum ${\rm FF}$ was obtained
were very scattered. After careful examination of the behavior of the
correction and the residual phases, we can confidently conclude that the
reason for this unexpected uncorrelation is the fact that (\ref{e:Apost})
can efficiently track {\em any kind} of bump in the phase. This
realization can also explain the very weak dependence of the results on
the specific form of the frequency dependence ( $f^{-2/3}$ or $f^{-1}$,
cf. Sec. VI of \cite{Apost96}): the bumps become wider as the frequency
increases and this effect can be reflected in any decreasing function of
$f$ multiplying ${\mathcal B}$ in (\ref{e:Apost}).

 \subsection{The procedure and some examples}\label{ss:procedure}
 
 In the previous sections we used specific examples to probe in detail the
main qualitative effects of precession on waveforms. The results of such
an analysis motivated us to develop a new family of templates, which
combined with the oscillatory form originally suggested by \cite{Apost96}
can increase the FF to desired levels. In what follows we summarize our
suggested complete procedure for searches of precessing inspiral signals,
It involves three successive steps:
 \begin{itemize}
 \item A non-precessing Newtonian search using the standard 2-parameters
chirp family \cite{SathyD91,DhuraS94,BalasD94}, where ${\mathcal M}_{\rm
max}$ and $t_{\rm max}$ that maximize ${\rm FF}$ are determined. The phase 
of the associated template $T_{\rm max}$ is
 \be
 \label{e:phase_newt}
 \phi_{\rm Newt} = 2\pi t_{\rm max}f + \frac{3}{128} \l(\pi {\mathcal
M}_{\rm max}\r)^{-5/3} f^{-5/3} + \phi_{\rm const}
 \ee
 where $\phi_{\rm const}$ is a constant phase. It is well known that the
maximization with respect to $\phi_{\rm const}$ is analytical, so that the
search is really only two-dimensional. However let us mention that this
phase can be computed and is used when plotting the various residual
phases.

We would like to mention 
that the maximization over $t_c$ can also be obtained, in a more 
computationally efficient way,
using FFT techniques. In particular, such 
algorithms are used for current searches with LIGO.
However, the version of the 
code used in this paper does not use this feature. Such an improvement 
of the code has now been implemented (after submission) 
and FFT will be used for future
work. We note that the first tests show an almost perfect
agreement between the FFT method and the one used for this paper ($< 1\%$),
and therefore it is not necessary to re-derive the results presented here 
and obtained by a grid-based method. However, the
use of the FFT technique will allow us to explore a greater parameter-space.

It is important to note that this step can be replaced by a PN search where the
two compact object masses can decouple. We have adopted a Newtonian for
reasons of computational efficiency, but we have shown (Paper I) that the
results don't change significantly as long that the order of the search in
this first level is consistent with the order used to construct the
non-precessing part of the signal. We will return to this issue in Sec.  
\ref{ss:results} (see also Paper I; cf. \cite{Apost96} for effects of
inconsistent choices).
 \item A three parameter search using the oscillatory correction
(\ref{e:Apost}) to the phase (\ref{e:phase_newt}). The values that maximize
FF are ${\mathcal B}_{\rm max}$, ${\mathcal C}_{\rm max}$, and
$\delta_{\rm max}$, and the phase of the newly determined template is
 \be
 \label{e:phase_mimic}
 \phi_0 = 2\pi t_{\rm max} f+ \frac{3}{128} \l(\pi {\mathcal M_{\rm
max}}\r)^{-5/3} f^{-5/3} + \phi_{\rm const} + {\mathcal C}_{\rm max} \cos
\l({\mathcal B}_{\rm max} f^{-2/3} + \delta_{\rm max}\r)
 \ee
 \item The last step consists of identifying the spikes in the residual
phase (\ref{e:phase_mimic}). The spikes are searched one by one, and for
each one a 2-parameter search is performed (plus the boolean parameter
$\varepsilon$). This sequential search ends when the relative change in
${\rm FF}$ is smaller than a given threshold. Let us call $N_{\rm max}$
the number of spikes found and $f_{0\,{\rm max}}\l(i\r)$, $\sigma_{\rm
max}\l(i\r)$, and $\varepsilon_{\rm max}\l(i\r)$ the frequency positions,
widths, and signs that maximize FF. The phase of the associated template
$T_{\rm max}$ is
 \bea
 \label{e:phase_spikes}
 \phi_{N_{\rm max}} = &2\pi t_{\rm max}f& + \frac{3}{128} \l(\pi {\mathcal
M_{\rm max}}\r)^{-5/3} f^{-5/3} + \phi_{\rm const} + {\mathcal C}_{\rm
max} \cos \l({\mathcal B}_{\rm max} f^{-2/3} + \delta_{\rm max}\r) \\
 \nonumber + &\sum_{i=1}^{N_{\rm max}}& P\l(f_{0\,{\rm max}}\l(i\r),
\sigma_{\rm max}\l(i\r), \varepsilon_{\rm max}\l(i\r)\r)
 \eea
 \end{itemize}
 We note that nothing at this point ensures that this step-wise procedure
produces the highest ${\rm FF}$ possible. We intend to perform tests and
maximizations over combined template parameters (4 and 5-dimensional) in
the near future, as our computational resources permit. However, we stress
that the procedure suggested at this point involves a number of
few-parameter, computationally efficient searches through template spaces.  
There is only one 3-dimensional search. In fact it seems that
four-parameter searches are beyond current and near-future developments in
computational resources \cite{VecchO02}.

\begin{figure}
 \includegraphics[height=10cm]{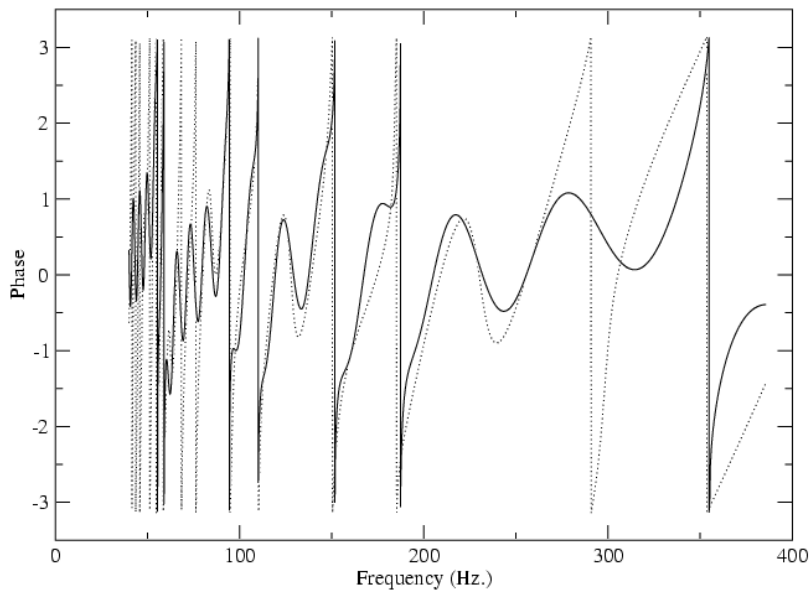}
 \caption{\label{f:seven} The dotted line is the residual phase $\Delta
\phi$ (Eq. (\ref{e:phires})), for the Configuration II (same as on the
right panel of Fig. \ref{f:oscille}). The solid line shows the ``mimic''
phase of the template, i.e., the sum of Eq. (\ref{e:Apost}) and seven
spikes (\ref{e:spike}). The resulting FF is equal to 0.79. It is clear
that the final ``mimic'' template matches the residual phase well. }
 \end{figure}

In what follows we give one example of how the above procedure can
increase the SNR and hence the detection rate. We use Configuration II
from Figure \ref{f:oscille}, which combines both oscillatory and spiky
behavior (second type of signal from the previous subsection).  Figure
\ref{f:seven} shows both the residual phase (i.e., the phase difference
between the signal and the Newtonian best-fitting template) and the
``mimic'' phase at the end of the search procedure.  In this particular
case, the value of ${\rm FF}$ converges to 0.79 after seven spikes have
been identified.

\begin{figure}
 \includegraphics[height=6.5cm]{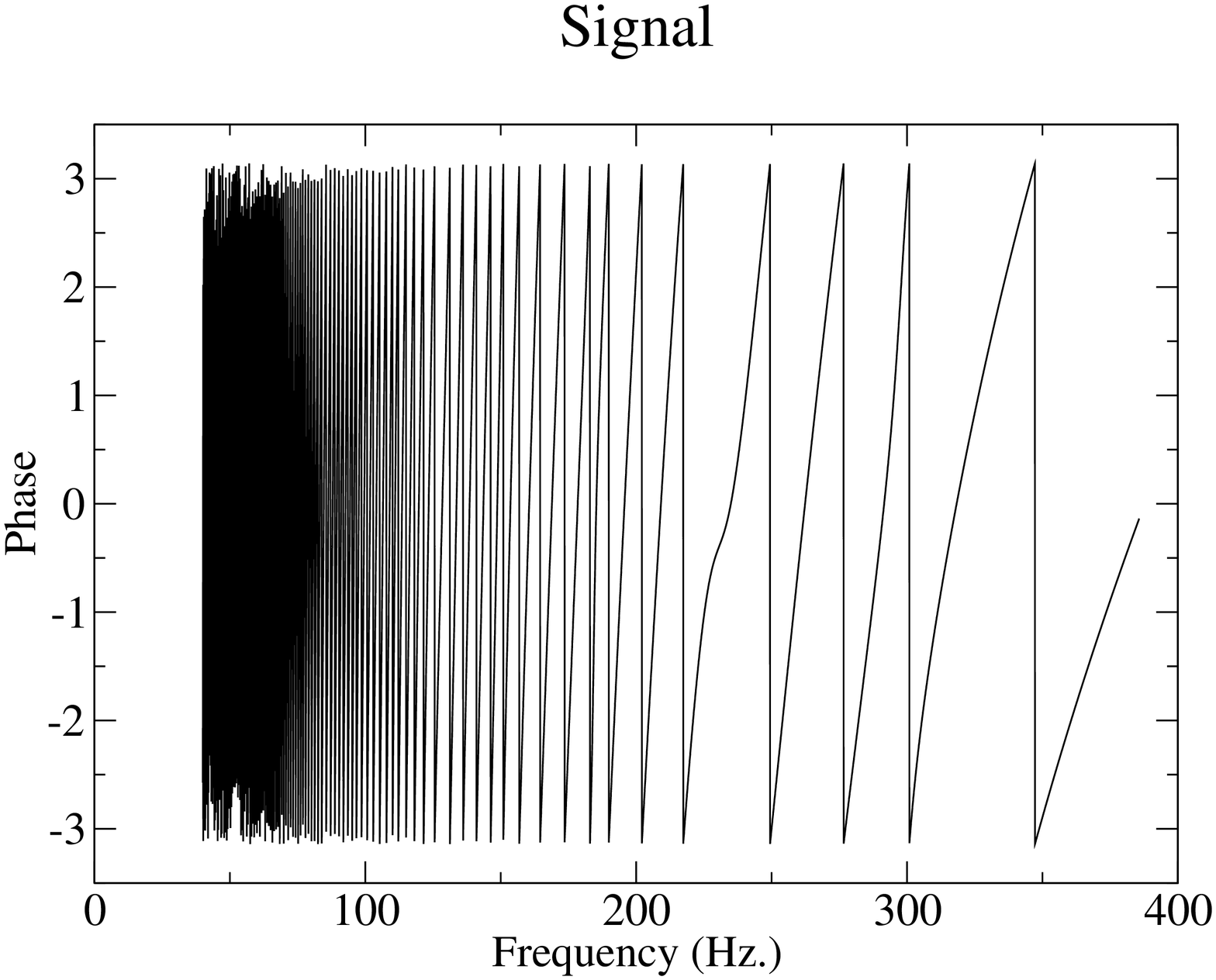}
 \includegraphics[height=6.5cm]{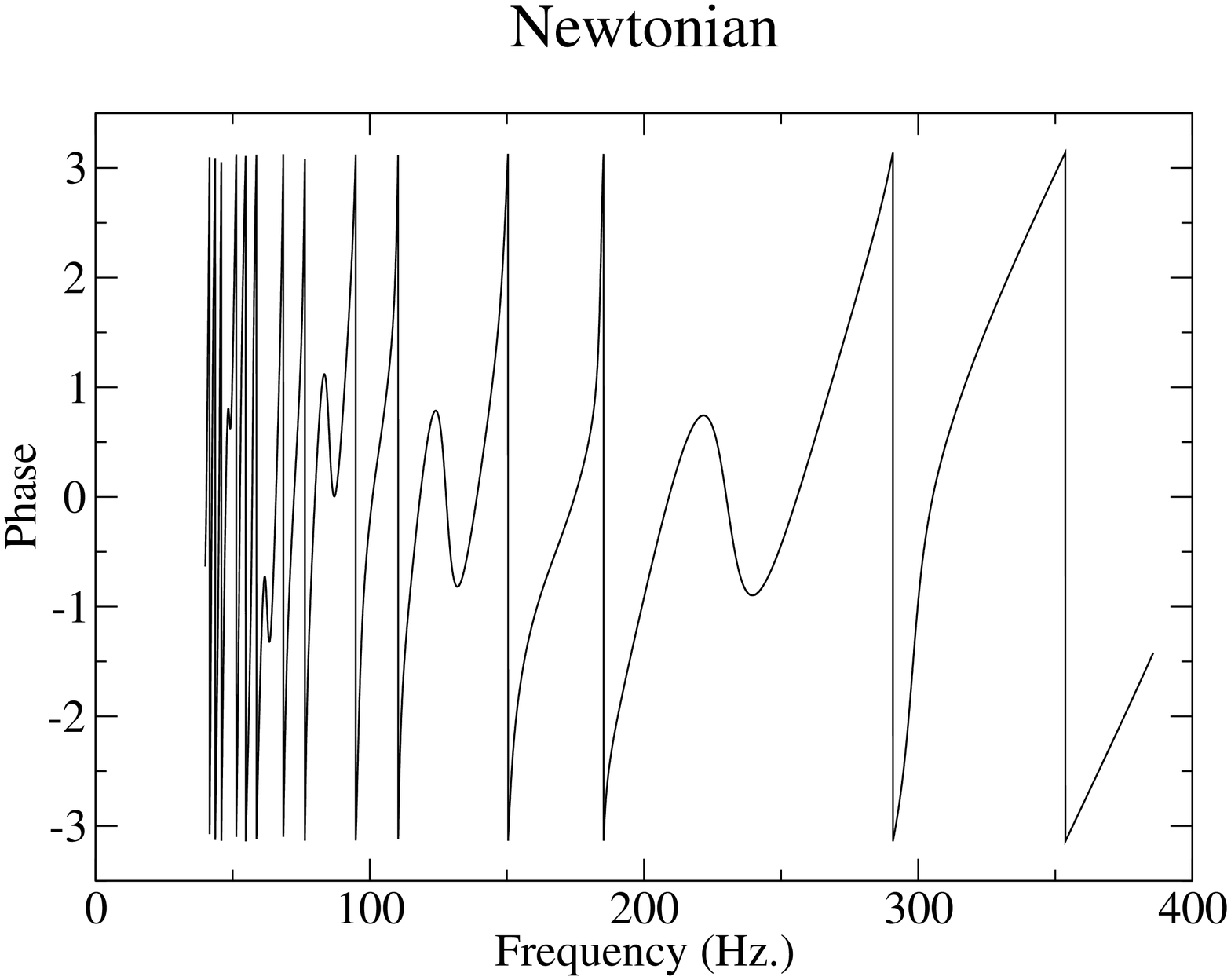}
 \includegraphics[height=6.5cm]{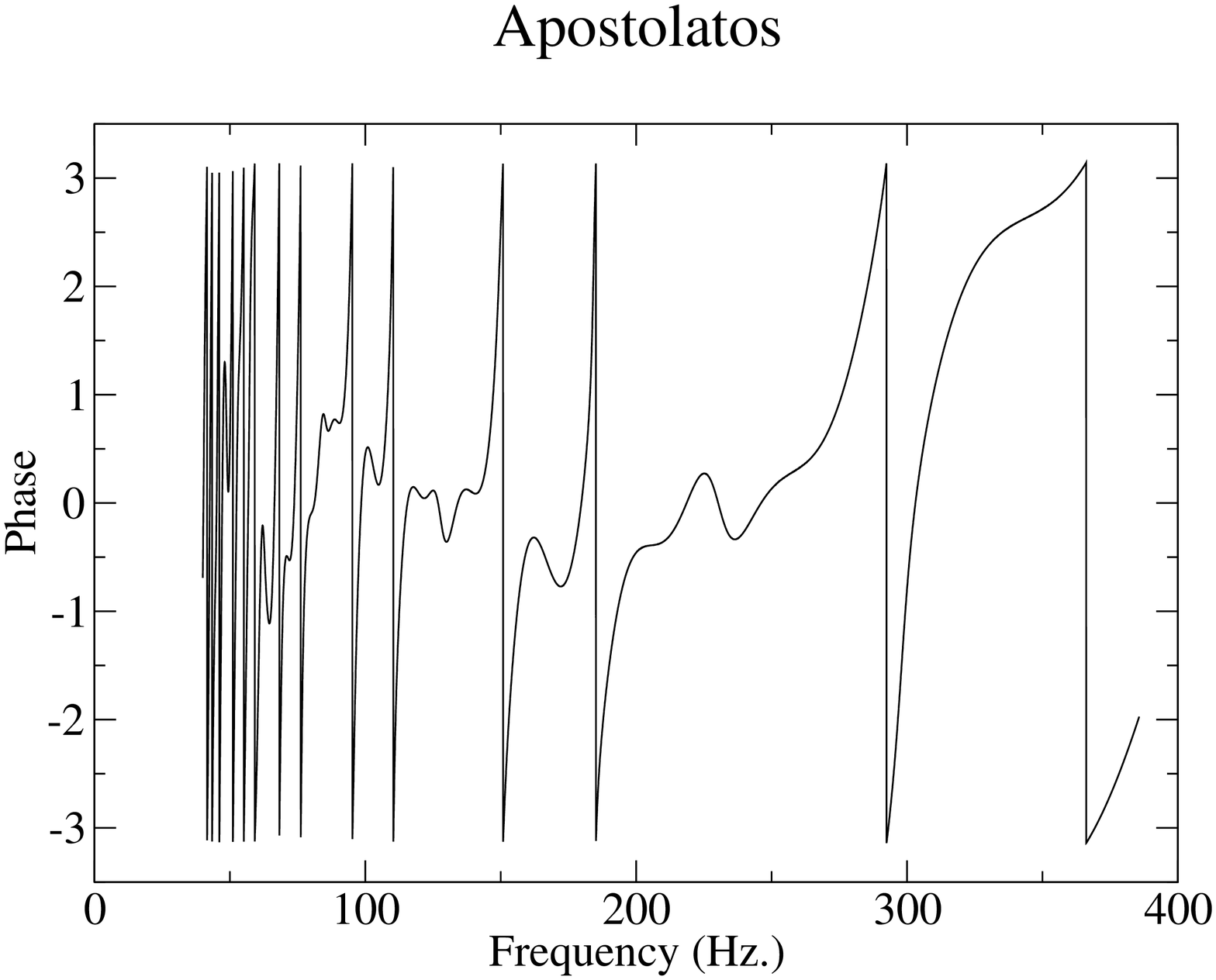}
 \includegraphics[height=6.5cm]{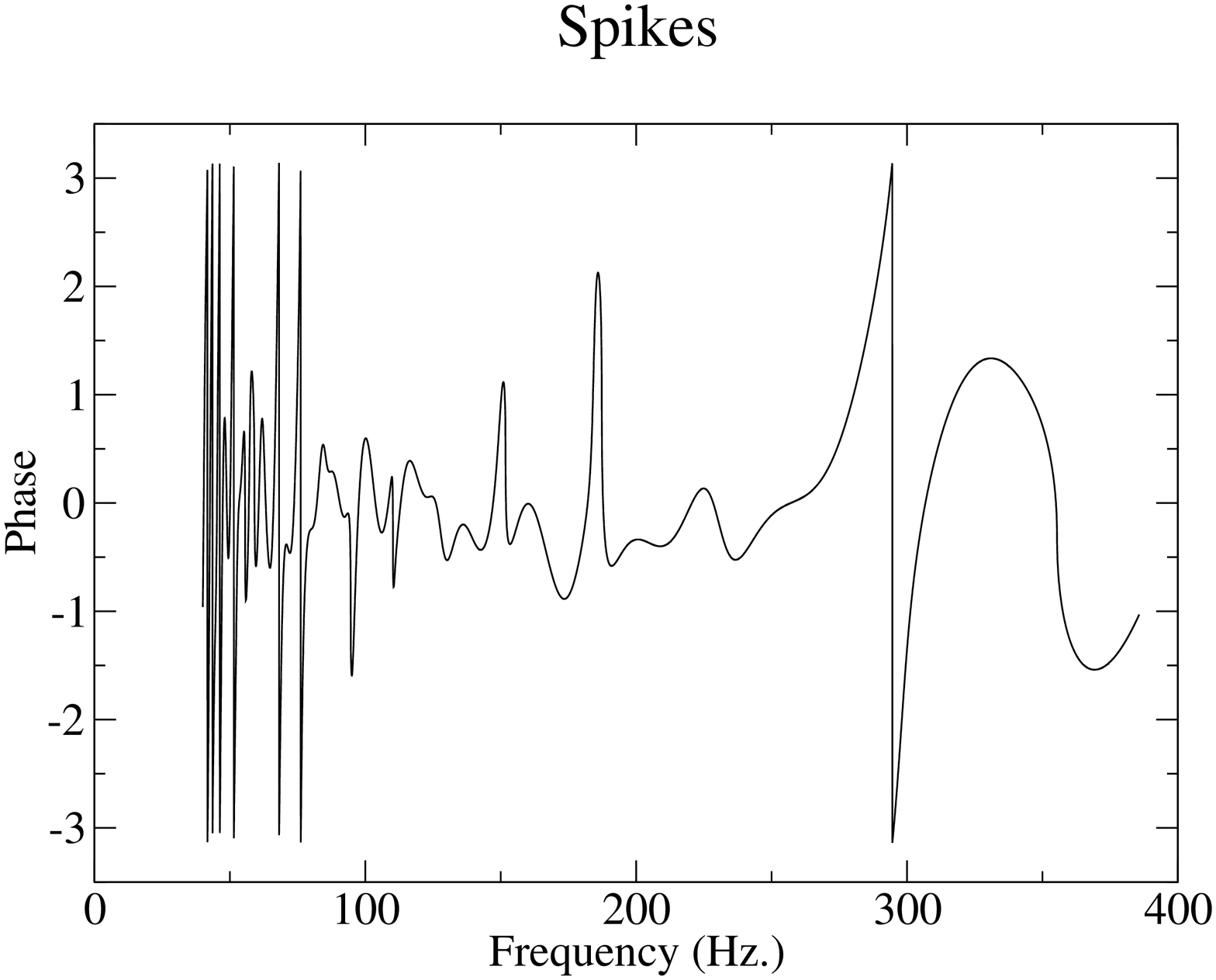}
 \caption{\label{f:substract} The top-left panel shows the phase of the
signal for the Configuration II. The other plots show the residual phase
after each step of the search procedure is completed: after Newtonian 
search (top right), oscillatory phase correction (bottom left), ``spiky'' 
phase correction (bottom right; seven spikes). } 
 \end{figure}

The plots of Fig. \ref{f:substract} provide another way of examining the
results. The first panel (top left) shows the phase of the signal
(configuration II in Table \ref{t:config}). The second one (top right)  
shows the residual phase between the signal and the best-fitting Newtonian
template, the third one (bottom left)  the residual phase after the best
fitting oscillatory form (\ref{e:Apost}) is incorporated in the template,
and the last one (bottom right)  the residual after all seven spikes have
been identified and included in the final template. Each step can be
viewed as an attempt to reduce the phase difference to zero, i.e. to make
the curves of Figure \ref{f:substract} as close to zero as possible.  
This particular example shows that our suggested procedure works very
well, especially in the regions of maximum sensitivity around 150 Hz.
Indeed the phase difference is drastically reduced in those regions. The
remaining spikes have not been found because they cause a very small
increase of the ${\rm FF}$, either because they are not very wide or
because they are located at frequencies where the noise is high.

 \begin{figure}
 \includegraphics[height=10cm]{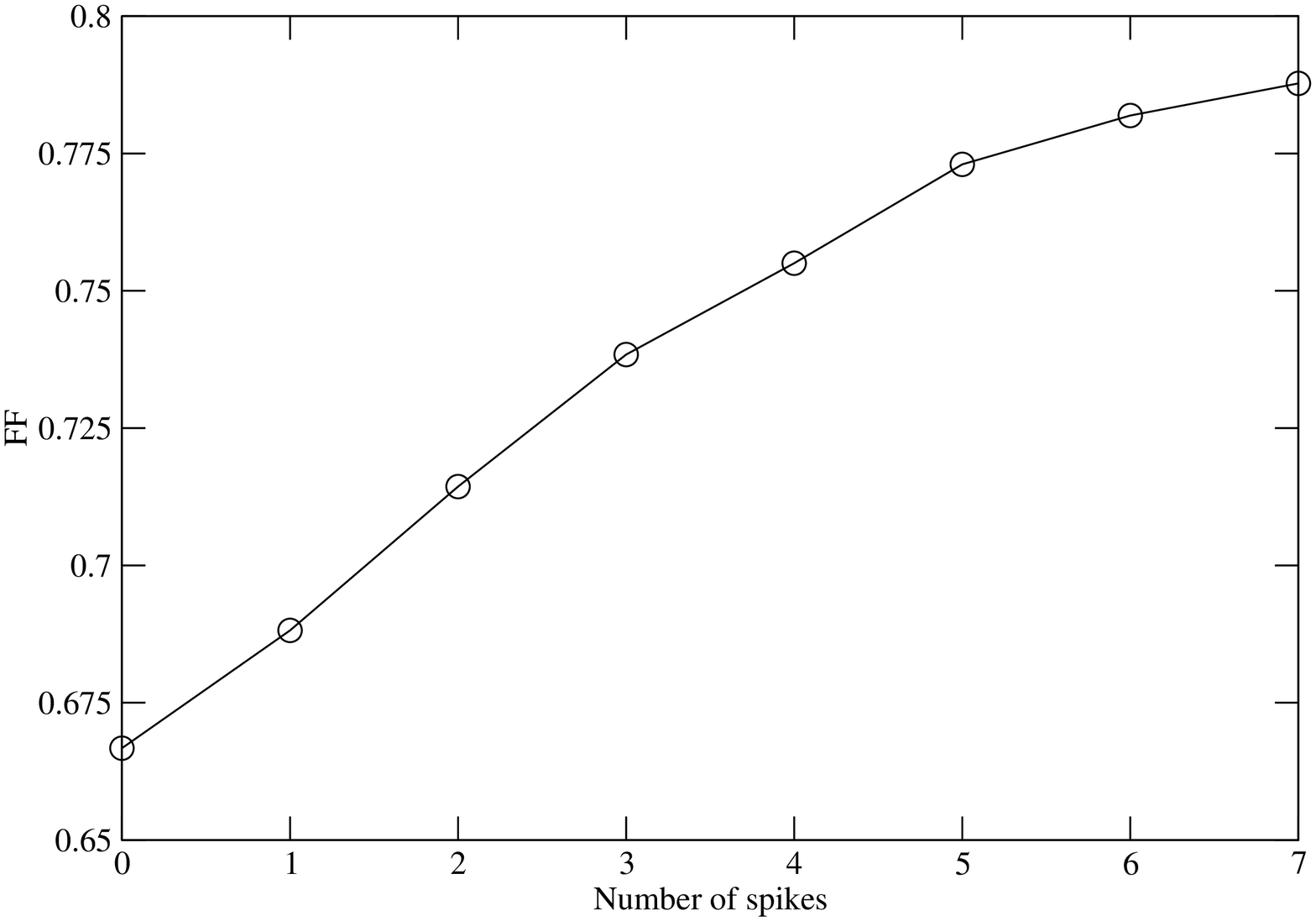}
 \caption{\label{f:conv_ff} Convergence of the ${\rm FF}$ value as more
and more spikes are added to the phase. The search was ended when the
relative change in ${\rm FF}$ dropped below $10^{-2}$ (Configuration II).}
 \end{figure}

Finally, in Figure \ref{f:conv_ff} we show the improvement in FF as more
and more spikes are identified and added to the best-fitting template. The
case $n=0$ corresponds to the case where only the oscillatory term
(\ref{e:Apost}) is included. The curve shows that, indeed, the more
efficient spikes (i.e. the ones giving rise to the best improvement in one
step), are found first. This is a clear and strong indication of the {\em
independence of the spikes}, which is crucial for the success of the
hierarchical search. The curve converges as less and less significant
spikes are identified and included. In this particular case, the ${\rm
FF}$ using just Newtonian templates is $0.55$. Both the oscillatory
correction and the spikes produce an improvement of about $20\%$. It is a
case where both types of corrections are important. This is actually
expected, as the residual phase (top right panel of Fig. \ref{f:oscille})
contains both bumps and spikes.

\begin{figure}
 \includegraphics[height=6.5cm]{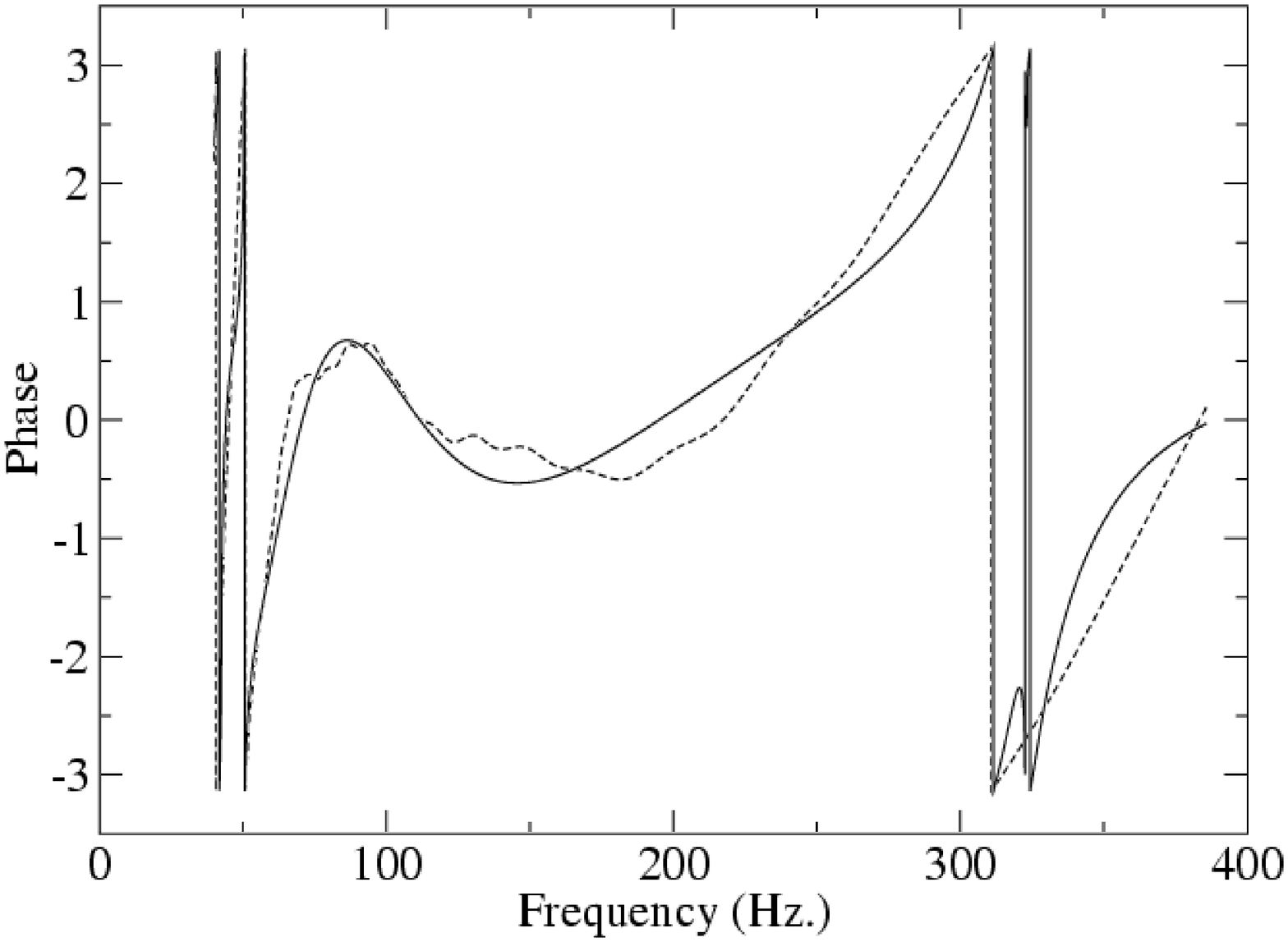}
 \includegraphics[height=6.5cm]{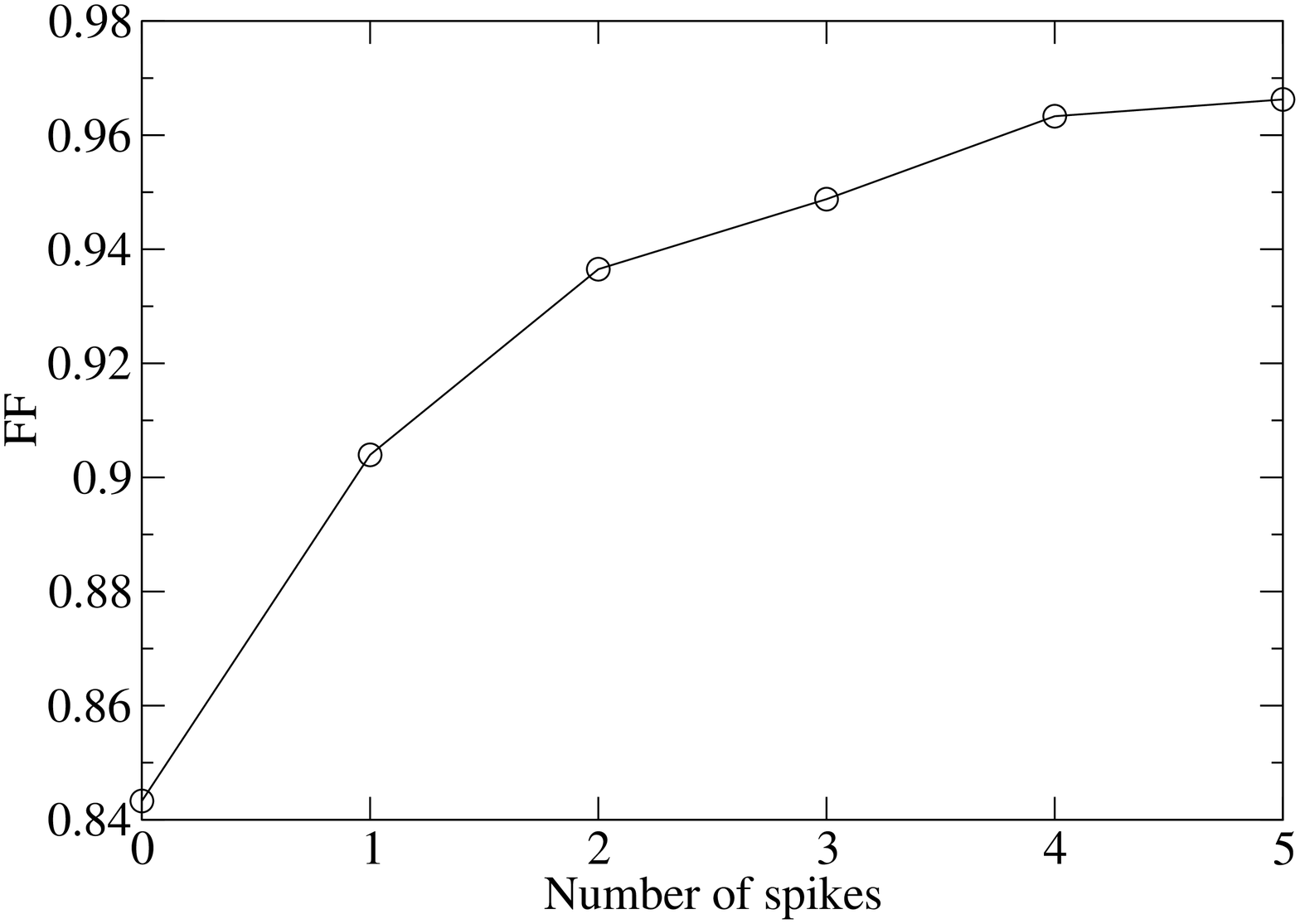}
 \caption{\label{f:best} The left panel shows the residual phase (dashed
line)  and its fit including an oscillatory phase correction and
five spikes (solid line). The right panel shows the convergence of the
${\rm FF}$ as more and more spikes are found. The physical parameters are
$m_1 = 10 \, M_\odot$, $m_2 = 1.4 \, M_\odot$, $S=1$ and $\kappa = 0.4$.
The random angles are given by $\cos \theta = -0.632$, $\varphi = 2.57$,
$\cos \theta' = 0.593$, $\varphi' = 5.65$ and $\alpha_{\rm prec} = 5.67$.}
 \end{figure}

Let us mention that the previous example is by far not one of the best
cases. In fact the final ${\rm FF}$ is somewhat smaller than the average
value. One of the best case (out of 2,000 sets of random angles) is shown
on Fig. \ref{f:best}. The left panel shows both the residual phase (dashed
line) and the fit (solid line) after the finding of five spikes.  The
convergence of the ${\rm FF}$ is plotted on the right panel and it shows
an impressive convergence to a value of almost $0.97$. The value using
just the Newtonian templates is $0.78$.

\subsection{Other possible corrections}\label{ss:amplitude}

 So far we have focused on an empirical approach that aims at reducing the
phase difference between the signal and the best possible template.
Indeed, matched-filtering techniques require that signal and template be
in phase for the longest possible frequency. Therefore one might expect
that phase corrections are more crucial than amplitude modulations. Once
the suggested procedure is completed, the residual phase appears to be
very noisy (e.g., bottom-right panel of Figure \ref{f:substract}). It
seems to us that it is almost impossible to find any additional systematic
correction. However, in an attempt to further increase the ${\rm FF}$
value, we examine whether a correction to the amplitude of 
the templates would be useful.

The amplitude modulation ${\rm AM}$, in the simple precession regime, is mainly
oscillatory (Eq. (11) of \cite{Apost95}). So
it is natural to introduce an amplitude correction with the same form as
Eq. (\ref{e:Apost}). Given that ${\rm FF}$ is independent of the absolute
normalization, we can drop the parameter ${\mathcal C}$ this time. Then we
are left with the following correction:
 \be
 \label{e:ampli}
 {\mathcal A}^{\rm cor} = \cos \l({\mathcal B}^{\rm ampli} f^{-2/3}
+\delta^{\rm ampli}\r).
 \ee

One must then determine the appropriate way to include such a correction.
One simple way is to apply it sequentially after all the corrections to
the phase (described above) have been applied. We have tested such an
approach and found almost no improvement at all in terms of increasing
${\rm FF}$. We conjecture that the mismatch of the non-precessing
parameters and all the modifications on the phase have already attempted
to ``mimic'' the amplitude modulation.  Alternatively we attempt to apply
the amplitude correction (\ref{e:ampli}) before any phase corrections
except for the Newtonian search. We decide to combine the amplitude
correction with the search with respect to the non-precessing parameters
in one single, {\em four-dimensional} maximization, by using the following
template
 \be
 \label{e:temple_ampli}
 \tilde{h}\l(f\r) = {\mathcal A}^{\rm cor} f^{-7/6} \exp\l(i \phi_{\rm
Newt}\r)
 \ee
 where $\phi_{\rm Newt}$ is given by Eq. (\ref{e:phase_newt}), so that the
template (\ref{e:temple_ampli}) is just the Newtonian template corrected
by an oscillatory term in the amplitude. Once this first step is
completed, we follow with the standard procedure of phase corrections,
starting with the oscillatory term and following with the spikes.

The first maximization mentioned above is four-dimensional, and hence
computationally demanding. Therefore we restrict it to a few sets of
random angles, typically ten. For each configuration, we calculate the
associated ${\rm FF}$ by using both the standard procedure (without any
amplitude corrections)  and the four-dimensional search (followed by the
oscillatory and ``spiky'' searches). Clearly exploring just 10 sets of
random angles is not adequate but it can still indicate whether the
amplitude correction is important or not.

 \begin{figure}
 \includegraphics[height=10cm]{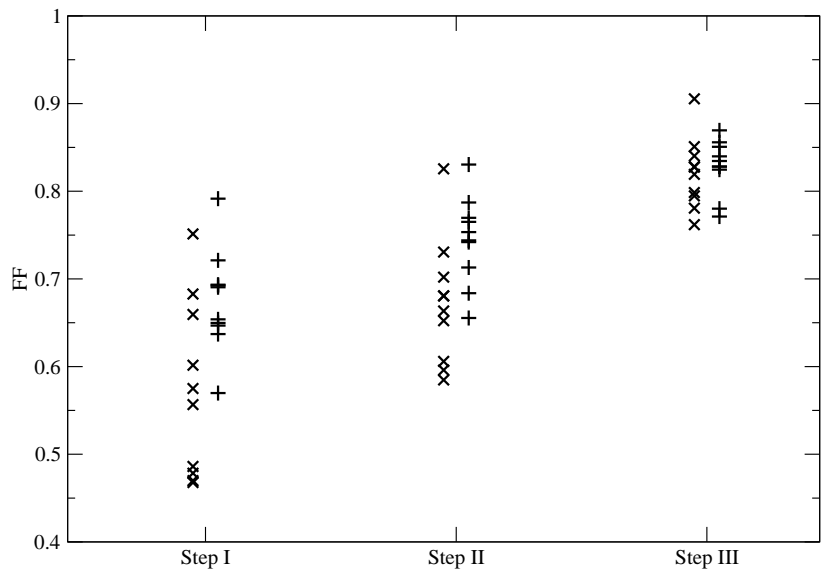}
 \caption{\label{f:steps}Comparison between the standard method ( $\times$
symbols) and the method including an amplitude correction ( $+$ symbols).
For $m_1 = 10$\,M$_\odot$, $m_2 = 1.4$\,M$_\odot$, $S=1$, and $\kappa =
0$, ten sets of random angles are shown (some of the points overlap).}
 \end{figure}

Figure \ref{f:steps} shows the comparison for ten values of the random
angles, for $m_1 = 10$\,M$_\odot$, $m_2 = 1.4$\,M$_\odot$, maximum spin
magnitude ($S=1$), and a spin-tilt angle of 90 degrees ($\kappa = 0$).  
Each symbol denotes the value of ${\rm FF}$ for one particular orientation
(some points are not visible due to overlap).  The $\times$ symbols
correspond to results for the standard procedure and the $+$ symbols to
those from the four-dimensional search. Between the two procedures, only
Step I is different; Step II is the search for oscillatory corrections
(\ref{e:Apost}), and Step III is the search for spikes.  We find that
initially the four-dimensional search leads to higher ${\rm FF}$ values in
comparison to the Newtonian and oscillatory search, but this effect is
overcome by the inclusion of spikes. At the end of the process, the
average fitting factor is almost the same for the two methods : $<{\rm
FF}> = 0.821$ for the standard procedure and $<{\rm FF}> = 0.828$ for the
case with the amplitude correction. Given these tests, we conclude that no
significant gain seems to be associated with the inclusion of an amplitude
modulation to the template family.

\begin{figure}
 \includegraphics[height=10cm]{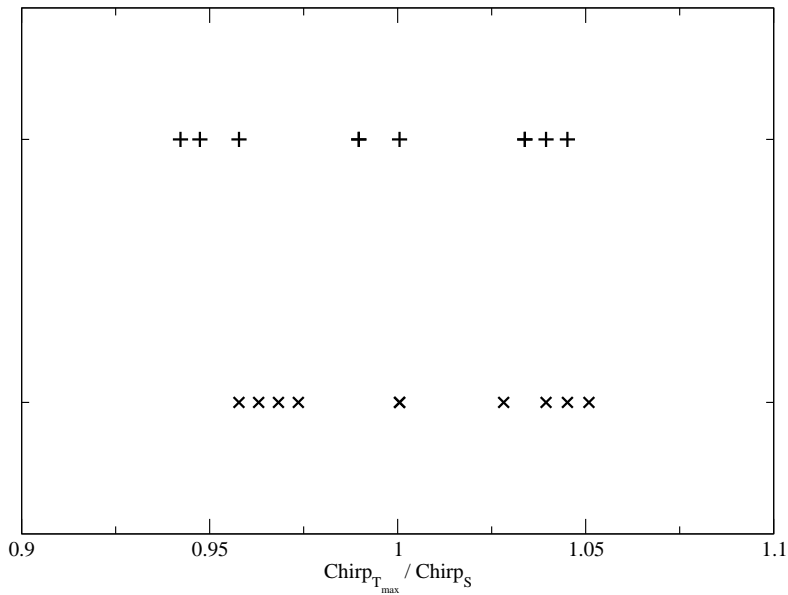}
 \caption{\label{f:disperse} Range of estimated chirp masses, measured by
the ratio of the estimated values (i.e. the ones of the template for which
the maximum ${\rm FF}$ is attained) to the ``true'' values (i.e. the ones
of the signal). The sets of random angles and the physical parameters are
the same than the ones of Fig. \ref{f:steps}. The $\times$ denote the
standard method and the $+$ the four-dimensional search.}
 \end{figure}

Even though $<{\rm FF}>$ values are very close, the procedure involving
the more computationally demanding four-dimensional search might still be
a viable choice, if, for example, the extraction of the physical
parameters of the binary is more accurate. To examine this question, we
compare the parameters of the signal (${\mathcal M}\simeq 2.99 M_\odot$
and $t_c=0$) to the best-fitting values obtained with the two different
methods. We focus on the chirp mass and show the results in Figure
\ref{f:disperse}. It is clear that the dispersion around the signal-value
of the chirp mass is very comparable for the two methods. We have
conducted tests for $m_1 = 7$\,M$_\odot$, $m_2=3$\,M$_\odot$, $S=1$, and
$\kappa = 0.4$ and observed very similar behaviors. Therefore, at this
point, it seems that there is no good reason for using the time-consuming
four-parameter search instead of the standard method described earlier.

\section{Efficiency of the ``spiky'' templates}\label{s:efficiency}

\subsection{Determination of the computational parameters}\label{ss:comput}

 To perform the searches we need to first determine a number of
computational parameters: the boundaries of the parameter space to be
searched (we do make use of our knowledge of the true parameters of the
signal), the number of templates between these boundaries, the number of
sets of random angles needed to obtain an accurate average $<{\rm FF}>$,
the number of collocation points needed to obtain an accurate value for
${\rm FF}$ with a Gauss-quadrature scheme used in the cross-correlation
calculations.

\begin{figure}
 \includegraphics[height=10cm]{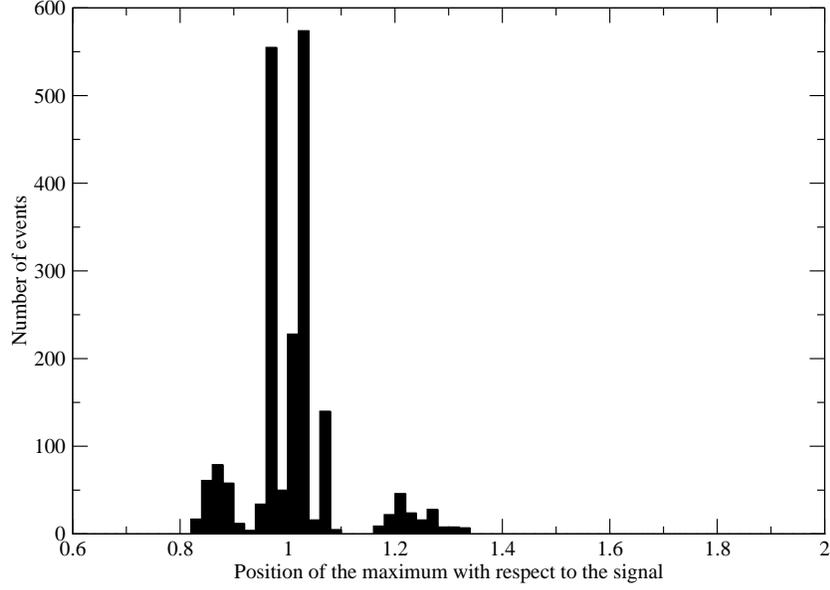}
 \caption{\label{f:pos_max} For 2,000 sets of random angles, the value of
${\mathcal M}_{\rm max}$
 at which the maximum of the ${\rm FF}$ is attained, is computed and the
results plotted in this histogram. The configuration is $m_1=10 M_\odot$,
$m_2 = 1.4 M_\odot$, $S=1$ and $\kappa = -0.5$. The range of the plot is
the actual range for the large search of the maximum (i.e.  $\l[0.6
{\mathcal M}_s, 2 {\mathcal M}_s\r]$)}
 \end{figure}

In what follows we briefly illustrate how the search intervals are found, 
using the chirp mass as an example. 
We first determine the
values ${\mathcal M}_{\rm min}$ and ${\mathcal M}_{\rm max}$, so that the
maximum value of the ${\rm FF}$ is always obtained for a value ${\mathcal
M}_{\rm max}$ inside this range.  Of course, for reasons related
to computational time, we would like to search through intervals as small
as possible. To determine the limits, for each choice of
masses, we consider one of the worst cases for the ${\rm FF}$, that is
$S=1$ and $\kappa = -0.5$ (see results of Paper I, for why this is the 
worst case). For this orientation, we use a very broad
interval for the search of the maximum.  Fig.\ \ref{f:pos_max} shows the
histogram of the occurrences at which the maximum of the ${\rm FF}$ is
obtained at a given ${\mathcal M}_{\rm max}$, for $m_1=10 M_\odot$, $m_2 =
1.4 M_\odot$, $S=1$ and $\kappa = -0.5$. In this particular, example the
search interval is then chosen to be $\l[0.8 {\mathcal M}_s, 1.3 {\mathcal
M}_s\r]$, where ${\mathcal M}_s$ is the chirp mass of the signal (less
than $1\%$ of the maxima lie outside this range, for the particular
realization of Fig. \ref{f:pos_max}).
 
\begin{figure}
 \includegraphics[height=10cm]{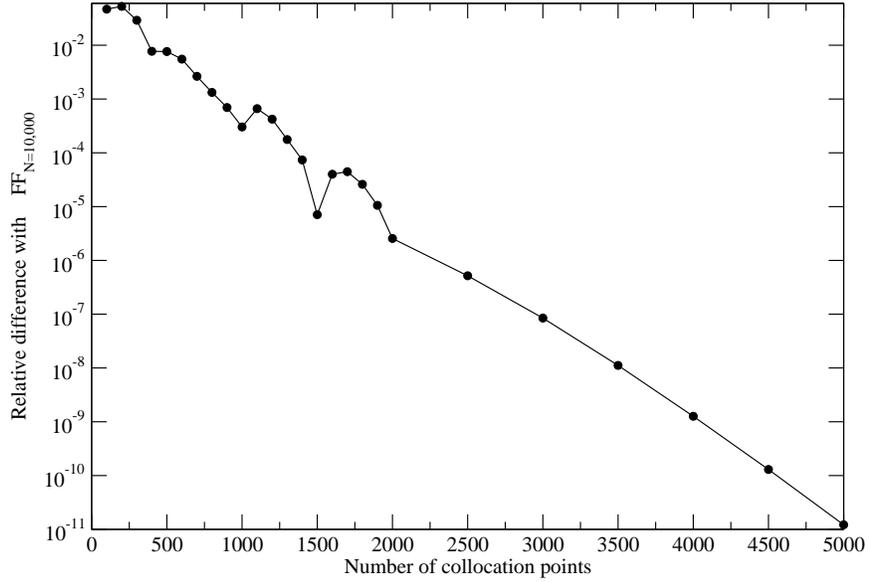}
 \caption{\label{f:coloc} Convergence of the ${\rm FF}$ with respect to
the number of collocation points. The signal is for $\l(10; 1.4\r)$ solar
masses objects, maximum spin and $\kappa = -0.2$. The random angles are
$\theta=0.107\pi$, $\varphi=0.292\pi$, $\theta'=0.412\pi$,
$\varphi'=1.85\pi$ and $\alpha_{\rm prec} = 0.499\pi$. The comparison is
done with the value for very high $N$ which is ${\rm FF} \simeq 0.46$. }
 \end{figure}

As we did in Paper I, the number of templates, sets of random angles, and
collocation points are determined by studying the relative convergence of
the ${\rm FF}$ as these numbers increase. Here we skip the details, but we
mention that the chosen numbers are sufficient to ensure an accuracy of
the ${\rm FF}$ calculation of 1\% or better. We also require that for
signals without any precession modulation we recover ${\rm FF} \geq 0.97$
Let us mention that we impose the convergence of ${\rm FF}$ for every
configuration and not of the average $<{\rm FF}>$, which is a stronger constraint.
This true for all computational parameters, except for the number of sets of random 
angles, for which only the convergence of the average $<{\rm FF}>$ makes sense.
For example, Fig.  \ref{f:coloc}
shows the relative convergence with the increasing number of collocation
points $N$ used to calculate the integrals in the expression of the ${\rm
FF}$. From this particular plot, we estimated that we should use 1,000
collocation points to get a relative error of order $10^{-3}$.
 
 The complete set of computational parameters used in our searches are
summarized in Table \ref{t:comput_newt}.  We also note that the grids
associated with the masses are in logarithmic scale to ensure proper
distribution of the templates.

 \begin{table}
 \caption{\label{t:comput_newt} Various computational parameters for the
(post)-Newtonian templates. The mass intervals are given in fraction of
the signal masses. ``Col'' is the number of collocation points and
``Angles'' the number of sets of random angles. Let us mention that the 
parameters related to $t_c$ have no meaning when FFT techniques are used.}
 \begin{tabular}{|c||c|c|c|c|c|c|c|c|}
 \hline
 Masses $\l(M_\odot\r)$ & Col & Angles & Interval ${\mathcal M}$ & Number
${\mathcal M}$ & Interval $t_c$ $\l({\rm s.}\r)$ & Number $t_c$ & Interval
$m_{\rm T}$ & Number $m_{\rm T}$\\
 \hline
 $\l(7;3\r)$ &$1,000$&$2,000$& $\l[0.9;1.15\r]$ & $90$ &
$\l[-0.05;0.05\r]$ & $70$ &
 $\l[1;2.4\r]$ & $20$\\
 $\l(10;1.4\r)$ &1,000& 2,000&$\l[0.8;1.3\r]$ & $90$ & $\l[-0.1;0.1\r]$ &
$130$ &
 $\l[0.5;2.6\r]$ & $30$\\
 \hline
 \end{tabular}
 \end{table}

The computational parameters for the addition of Apostolatos' correction
are given in Tab. \ref{t:comput_mimic} and the ones for the spikes in
Table \ref{t:comput_spikes}. For each spike, $f_0$ is searched in all the
range of frequency, from the low-cut at 40 Hz to the termination frequency
at the ISCO. The value of the ISCO is the one of a test particle orbiting
a Schwarzschild black hole. It may seem a rather crude approximation but,
because the noise is high at those frequencies, its exact value is not
expected to be important.  A logarithmic grid is used for $f_0$, because
the spikes are more densely distributed at low frequencies. We search for
spikes until the relative change of fitting factor is smaller than $\delta
{\rm FF} = 10^{-2}$.

 \begin{table}
 \caption{\label{t:comput_mimic} Various computational parameters for the
mimic templates.}
 \begin{tabular}{|c||c|c|c|c|c|c|}
 \hline
 Masses $\l(M_\odot\r)$ & Interval ${\mathcal B}$ & Number ${\mathcal B}$
& Interval ${\mathcal C}$ & Number ${\mathcal C}$ & Interval $\delta$ &
Number $\delta$\\
 \hline

$\l(7,3\r)$ & $\l[0, 75\r]$ & 30 &$\l[0,\pi\r]$&10&$\l[0,2\pi\r[$&20\\
$\l(10,1.4\r)$ & $\l[0, 160\r]$ & 580
&$\l[0,\pi\r]$&10&$\l[0,2\pi\r[$&10\\
 \hline
 \end{tabular}
 \end{table}

 \begin{table}
 \caption{\label{t:comput_spikes} Computational parameters for the search
of the spikes.}
 \begin{tabular}{|c||c|c|c|c|c|}
 \hline Masses $\l(M_\odot\r)$ & Interval $f_0$ & Number $f_0$ & Interval
$\sigma$ & Number $\sigma$ & $\delta {\rm FF}$ \\
 \hline $\l(7 ; 3\r)$ & $\l[40 ; 440\r]$ & $500$ & $\l]0 ; 0.8\r]$ & 100 &
$10^{-2}$ \\ $\l(10 ; 1.4\r)$ &$\l[40 ; 386\r]$ & 400 & $\l]0 ; 0.8\r]$ &
60 & $10^{-2}$\\
 \hline
 \end{tabular}
 \end{table}

\subsection{Results}\label{ss:results}

\begin{figure}
 \includegraphics[height=6.5cm]{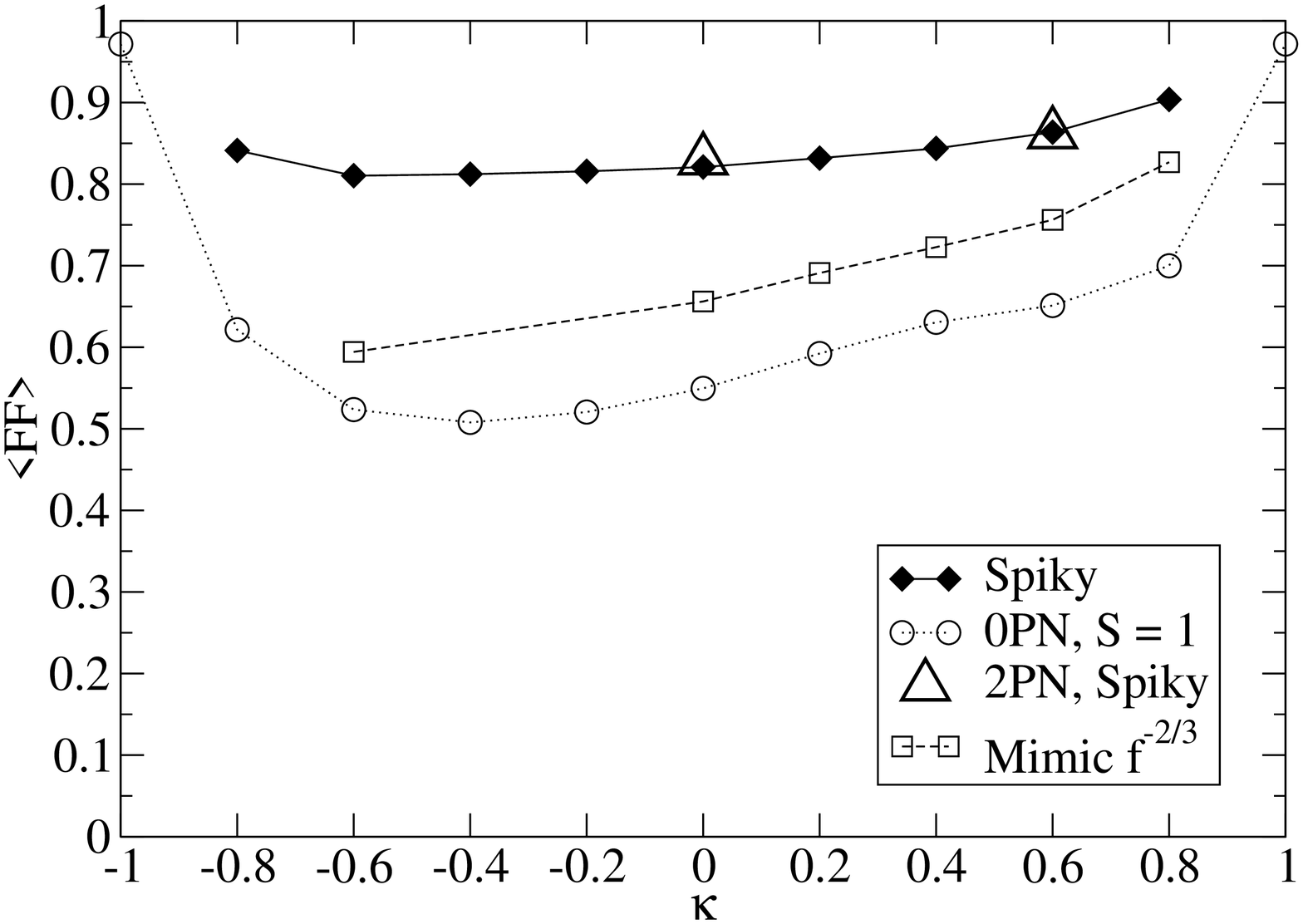} 
 \includegraphics[height=6.5cm]{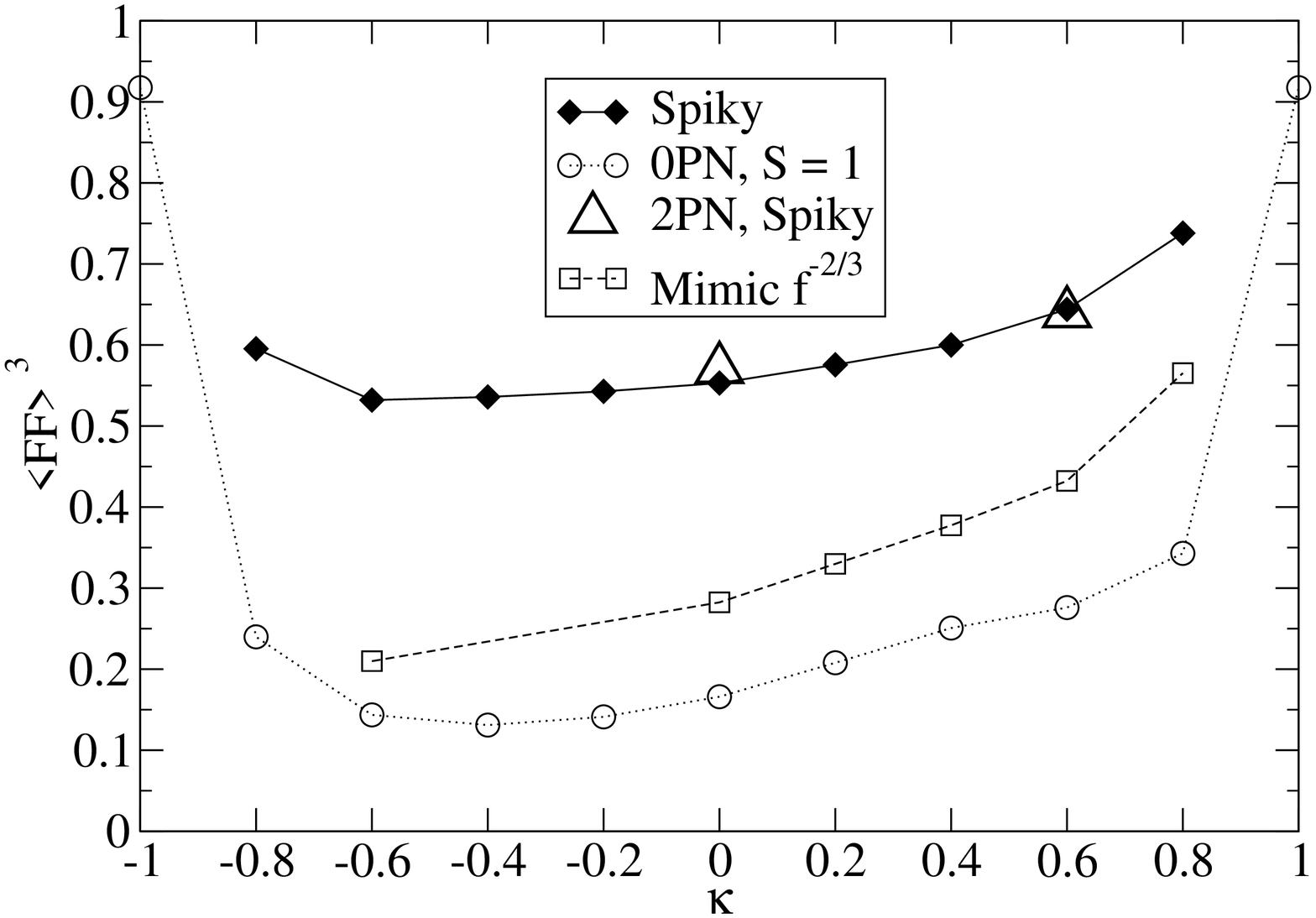}
 \caption{\label{f:spikes_10_14} Efficiency of the ``spiky'' templates in
recovering the signal-to-noise ratio and increasing the inspiral detection
rate.  The circles and dotted lines correspond to the values obtained
using the Newtonian templates. The squares and dashed lines correspond to
Apostolatos' waveforms alone, and the filled diamonds and solid lines
correspond to the combination of the Apostolatos' and ``spiky'' templates.
The triangles denote the values obtained by using 2PN order expressions
for the non-precessing part for {\em both} the signal and the templates.  
The left panel shows the average $<{\rm FF}>$ whereas the right panel
shows the reduction factors of the detection rate, both as a function of
the cosine of the spin-misalignment angle.}
 \end{figure}

 In Figure \ref{f:spikes_10_14} we show the efficiency of the ``spiky''
templates for $m_1 = 10$\,M$_\odot$, $m_2 = 1.4$\,M$_\odot$, and $S=1$.  
The left panel shows the average $<{\rm FF}>$ as a function of the cosine
of the misalignment angle $\kappa \equiv \vec{S}\cdot\vec{L}$ and the
right panel the reduction factor in detection rate $<{\rm FF}> ^3$,
assuming a uniform volume distribution of sources.  It is evident that the
new family of template can greatly improve the signal-to-noise ratio.  
Values of $<{\rm FF}>$ are above 0.8 for the full range of
spin-misalignment angles. Consequently the reduction factors of detection
rate increase by factors of up to 5-6 (compared to cases where precession
is ignored; see Paper I) or in other words the detection rate is never
reduced by more than a factor of 2.  This is to be compared with the
curves showing the Newtonian and Apostolatos' templates for which the
detection rate can be reduced by as much as 10 and 5 (respectively).

The triangles are computed by including 2PN corrections to the phases of
the non-precessing parts of {\em both} the signal and the templates. As
already stated in Sec. \ref{ss:procedure}, the results are almost the same
as the ones obtained by using only Newtonian expressions. It is a clear
validation of our results, as it illustrates the fact, already observed in
Paper I, that the order of the non-precessing part is not important, as
long as the order of the non-precessing parts is kept consistent between
the signal and the templates.

\begin{figure}
 \includegraphics[height=6.5cm]{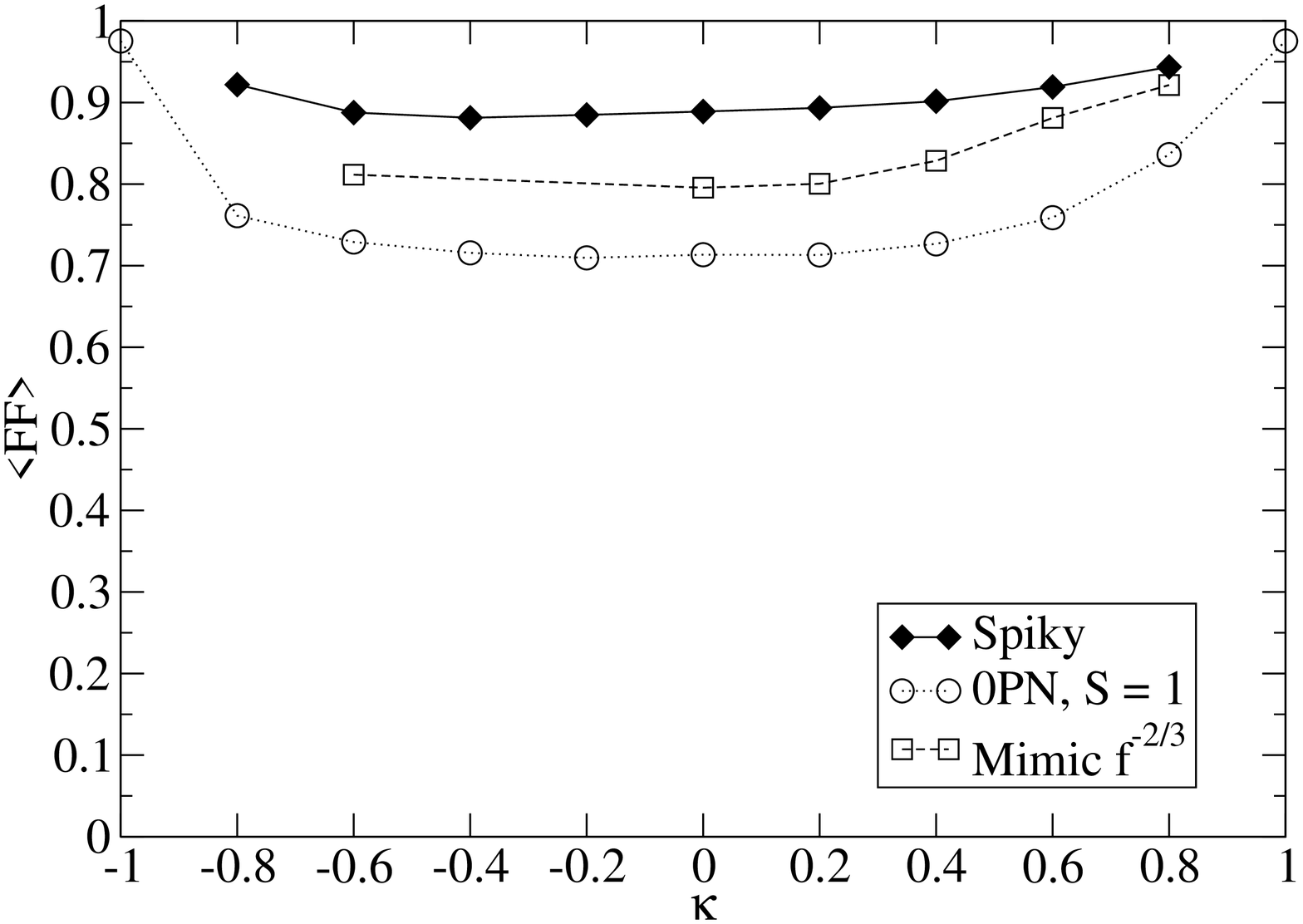} 
 \includegraphics[height=6.5cm]{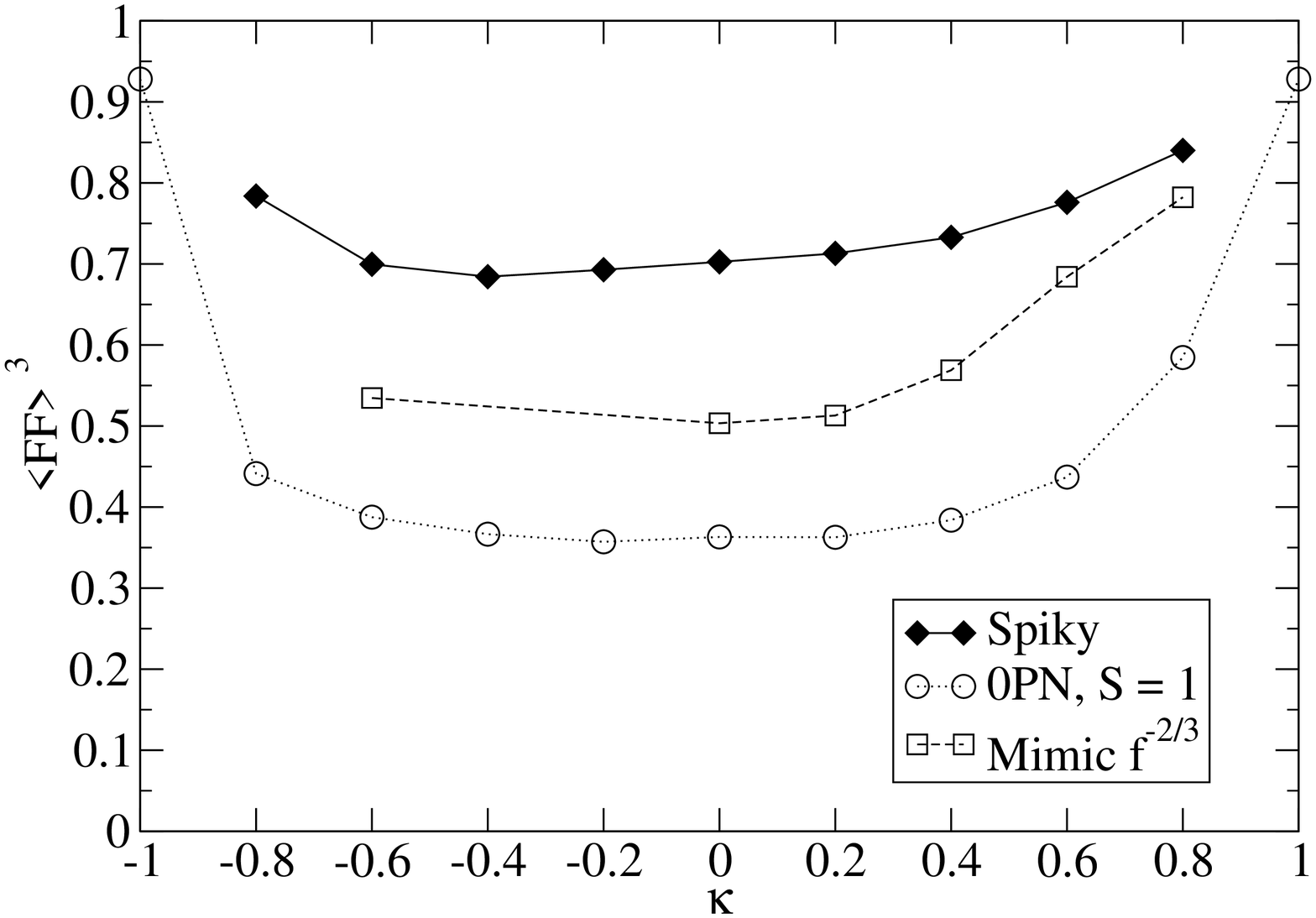}
 \caption{\label{f:spikes_7_3} Same as Fig. \ref{f:spikes_10_14} but for
$m_1 = 7 \, M_\odot$ and $m_2 = 3\, M_\odot$ (except for the 2PN
triangles).}
 \end{figure}

Figure \ref{f:spikes_7_3} shows the same results but for $m_1 = 7\,
M_\odot$ and $m_2 = 3\, M_\odot$. The reduced detection rate $<{\rm FF}>^3$ 
is always greater than 70\%, whereas it could drop to 50\% if one uses only
Apostolatos' correction.

\begin{figure}
 \includegraphics[height=10cm]{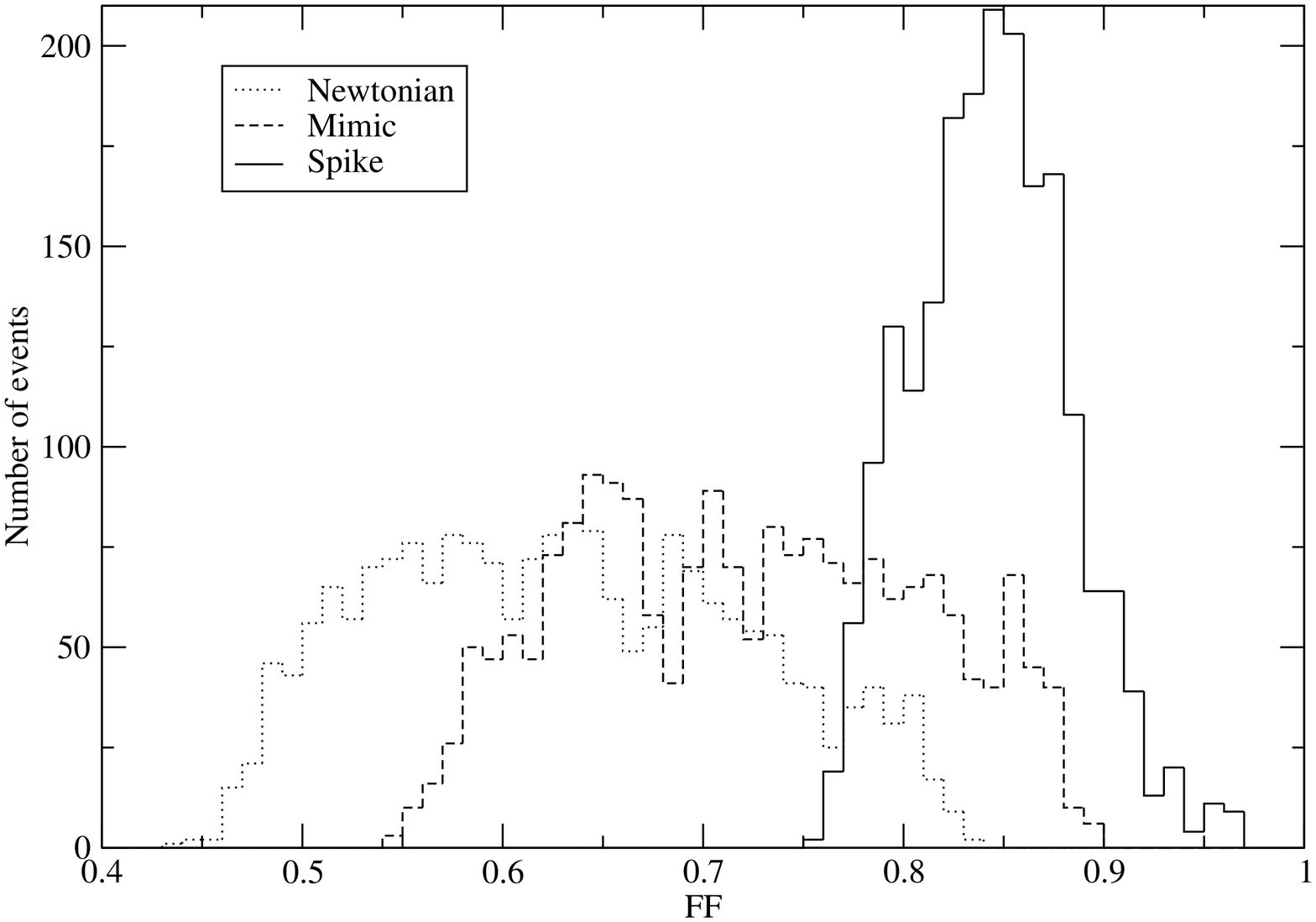} 
 \caption{\label{f:histo} Distribution of ${\rm FF}$
values using 2,000 configurations (sets of random angles). The physical
parameters adopted for the signal are $m_1=10 M_\odot$, $m_2 = 1.4
M_\odot$, $S=1$ and $\kappa = 0.4$. The dotted line denotes the values
after the use of the Newtonian templates, the dashed line those obtained
after the addition of the Apostolatos' oscillatory correction, and the
solid curve those after all the important spikes have been identified.}
 \end{figure}

Figure \ref{f:histo} presents another way of looking at the improvement
caused by the ``spiky'' templates. For $m_1 = 10 \, M_\odot$, $m_2 = 1.4
\, M_\odot$, $S=1$ and $\kappa =0.4$, the ${\rm FF}$ for 2,000 sets of random
angles has been computed. Figure \ref{f:histo} shows the associated
histogram, i.e. the number of configurations for which a given ${\rm FF}$
is obtained. One can clearly see that, after each step of our search
procedure, (i) the $<{\rm FF}>$ increases and (ii) the distribution of
${\rm FF}$ values narrows significantly.

As already stated in Sec. \ref{ss:procedure}, FFT techniques can also be used 
to compute the $<{\rm FF}>$ during the first step of our procedure
(post-Newtonian search). Although such algorithms have not been used for 
obtaining the plots shown in this paper, we conducted a few tests to check 
the validity of our results. As expected, we found that the agreement 
between the two methods is almost perfect. Indeed, it is better than a fraction
of $1\%$ for reasonable number of points used to compute the FFT.

\section{Summary and Conclusions}\label{s:conclu}

 The goal of this paper has been to find a new family of templates for the
efficient detection of precessing binary inspiral. We are motivated by a
careful study of the qualitative behavior of the residual phase once the
signal has been matched with a best-fitting non-precessing template. This
initial study provides with an explanation for the insufficiency of the
purely oscillatory phase correction shown in Eq. (\ref{e:Apost}),
suggested for the first time by Apostolatos \cite{Apost96}. The problem
seems to be that the mismatch between the non-precessing parameters of the
signal and the templates fails to remove all the monotonic behavior of the
phase modulation (cf. Fig. \ref{f:mono}). In some other cases, the
mismatch can even ``pollute'' the phase modulation and becomes responsible
for the appearance of a monotonic component (cf. Fig. \ref{f:oscille}).

To tackle this problem, we have shown that the monotonic behavior of the
phase can be represented by a succession of spike-like structures. These
spikes can be searched very efficiently by using a hierarchical algorithm,
in which the spikes are found one after the other until the ${\rm FF}$
converges. We note that, before searching for the spikes, it is useful to
still use the oscillatory correction \cite{Apost96} because the oscillator
(\ref{e:Apost}) can account for any periodic behavior present in the
residual phase. We also considered including a correction to the
amplitude, in the form of an oscillator similar to Eq. (\ref{e:Apost}). 
Our preliminary examination of this question did not reveal any 
significant improvement of the signal-to-noise ratio and therefor this 
approach did not warrant any further exploration. 

Quantitative results about the usefulness of the ``spiky'' templates in
improving detection rates have been presented, studying two configurations
for which precession is most important ($\l(m_1 , m_2\r) = \l(10 , 1.4\r)$
and $\l(7 , 3\r)$ M$_\odot$). We find that, even in the most unfavorable
cases, the detection rate is reduced by less than a factor of 2.
This is to be compared with the factor of 10
reduction when precession is ignored altogether and the factor of 5 when
the Apostolatos' oscillatory correction is included. 

We view our results as very encouraging, especially when one considers the
very moderate number of parameters involved (two for each spike). Of
course, it must be possible to obtain higher ${\rm FF}$ by using templates
with more parameters, but the applicability to real searches may be
problematic in terms of the associated computational efficiency. In our
effort to find this new family of templates, keeping the number of
parameters as low as possible was one of the main priorities, as long as
the improvement in the detection rate is satisfactory.

In this study we focused on trying to quantify the efficiency of the newly
proposed template family in increasing the SNR.  For the immediate future
there are a number of issues that we would like to examine in more detail,
as our computational resources permit. The ultimate question is whether
the ``spiky'' family of templates and the search procedure described here
can be used in real searches.

As the very first step, we would like to examine this family's efficiency
when we move away from the approximations of simple precession and we
consider more complicated precession signals, obtained by numerical
integration of more complete post-Newtonian dynamics for example
\cite{Kidde95}. We expect that the results will not change dramatically
because of the generic nature of our family of templates. Indeed, the
procedure presented in this paper should reproduce rather accurately any
superposition of oscillatory (Apostolatos' ansatz) and monotonic (the
spikes) behaviors in the phase. Moreover, we found that the results did
not change very much when higher order PN-effects were included in the
non-precessing part (cf. Sec.  \ref{ss:results}). Low false alarm rates
and moderate numbers of templates needed for real searches are amongst the
other properties that must be asserted. Issues related to reliable
physical parameter estimation are also important, if we want to think
about questions of interest to gravitational-wave astrophysics that goes
beyond detection. We plan to conduct these studies in the near future and
implement the ``spiky'' templates in LIGO algorithm library (LAL)
\cite{LAL}, should they constitute a viable family.

 \begin{acknowledgments}
 This work is supported by NSF Grant PHY-0121420. VK also acknowledges
support from a Science and Engineering Fellowship by the David and Lucile
Packard Foundation. We are also grateful to the High Energy Physics Group
at Northwestern University for allowing us access to their computer
cluster THEMIS. 
 \end{acknowledgments}


\begin{thebibliography}{}

\bibitem{GrandKV02} P. Grandcl\'ement, V. Kalogera and A. Vecchio, Phys.
Rev. D, submitted, gr-qc/0207062 (2002) (Paper I). 

\bibitem{Apost96} T.A.~Apostolatos, Phys. Rev. D {\bf 54}, 2421 (1996).

\bibitem{Abram92} A.~Abramovici {\em et al.}, Science {\bf 256}, 325
(1992).

\bibitem{Caron97} B.~Caron {\em et al.}, Nucl. Phys. {\bf B54}, 167
(1997).

\bibitem{Danzm95} K.~Danzmann, in {\em Gravitational Waves Experiments},
eds.  E.~Coccia, G.~Pizzella and F.~Ronga, World Scientific, Singapore
(1995).

\bibitem{Tagos01} H.~Tagoshi {\em et al.}, Phys. Rev. D. {\bf 63}, 062001
(2001).

\bibitem{Helst68} C.W.~Helstrom, {\em Statistical Theory of Signal
Detection}, 2nd edition, Pergamon Press, London (1968).

\bibitem{OwenS99} B.J.~Owen and B.S.~Sathyaprakash, Phys. Rev. D {\bf 60},
022002 (1999).

\bibitem{Finn99} L.S.~Finn, Written version of lectures given at XXVI SLAC
Summer Institute on Particle Physics {\em Gravity: From the Hubble Length
to the Planck Length}, gr-qc/9903107, (1998).

\bibitem{CutleT02} C.~Cutler and K.~S.~Thorne, {\em An Overview of
Gravitational-Wave Sources}, to appear in Proceedings of GR16 (Durban,
South Africa, 2001), gr-qc/0204090.

\bibitem{BradyCT98} P.R.~Brady, J.D.E.~Creighton and K.S.~Thorne, Phys.
Rev. D {\bf 58}, 061501 (1998).

\bibitem{BuonaCV02} A.~Buonanno, Y.~Chen and M.~Vallisneri, Phys. Rev. D
submitted, gr-qc/0205122 (2002).

\bibitem{ApostCST94} T.A.~Apostolatos, C.~Cutler, G.J.~Sussman and
K.S.~Thorne, Phys. Rev. D {\bf 49}, 6274 (1994).

\bibitem{Apost95} T.A.~Apostolatos, Phys. Rev. D {\bf 52}, 605 (1995).

\bibitem{SathyD91} B.S.~Sathyaprakash and S.V.~Dhurandhar, Phys. Rev. D
{\bf 44}, 3819 (1991).

\bibitem{DhuraS94} S.V.~Dhurandhar and B.S.~Sathyaprakash, Phys. Rev. D
{\bf 49}, 1707 (1994).

\bibitem{BalasD94} R.~Balasubramanian and S.V.~Dhurandhar, Phys. Rev. D
{\bf 50}, 6080 (1994).

\bibitem{VecchO02} B.~Owen and A.~Vecchio, in progress.

\bibitem{Kidde95} L.E.~Kidder, Phys. Rev. D {\bf 52}, 821 (1995).

\bibitem{LAL} LIGO/LSC Algorithm Library Home Page :
	http://www.lsc-group.phys.uwm.edu/lal/ 

\end{thebibliography}
\end{document}